\documentclass[pra,twocolumn,preprintnumbers,amsmath,amssymb,floatfix,superscriptaddress,showpacs]{revtex4}
\usepackage{graphicx}
\usepackage{graphics}
\usepackage{psfrag}
\usepackage{hyperref}
\usepackage{multirow}
\usepackage[sort&compress]{natbib}

\newcommand{\rmi}{{\rm i}}

\begin{document}

\hypersetup{pdftitle={Dimensional crossover of nonrelativistic bosons}}
\title{Dimensional crossover of nonrelativistic bosons}
\date{\today}
\author{Soeren Lammers}
\affiliation{Institute for Theoretical Physics, University of Heidelberg, D-69120 Heidelberg, Germany}
\author{Igor Boettcher}
\affiliation{Department of Physics, Simon Fraser University, Burnaby, British Columbia V5A 1S6, Canada}
\author{Christof Wetterich}
\affiliation{Institute for Theoretical Physics, University of Heidelberg, D-69120 Heidelberg, Germany}

\begin{abstract}
We investigate how confining a transverse spatial dimension influences the few- and many-body properties of non-relativistic bosons with pointlike interactions. Our main focus is on the dimensional crossover from three to two dimensions, which is of relevance for ultracold atom experiments. Using Functional Renormalization Group equations and T-matrix calculations we study how the phase transition temperature changes as a function of the spatial extent of the transverse dimension and relate the 3D and 2D s-wave scattering lengths. The analysis reveals how the properties of the lower-dimensional system are inherited from the higher-dimensional one during the renormalization group evolution. We limit the discussion to confinements in a potential well with periodic boundary conditions and argue why this qualitatively captures the physics of other compactifications such as transverse harmonic confinement as in cold atom experiments.
\end{abstract}


\maketitle


\section{Introduction}\label{SecIntro}

Lower-dimensional systems play a prominent role in statistical and condensed matter physics as they exhibit several unusual features due to the pronounced influence of fluctuations. At the same time, many technologically interesting materials such as high-temperature superconductors or graphene are based on layered or striped crystalline structures and thus inherit part of their properties from the reduced dimensionality. To disentangle the influence of dimensionality effects from other aspects of the many-body system constitutes a key challenge in advancing our understanding of condensed matter and in devising new promising materials. 

With the recent progress in trapping ultracold quantum gases it has become feasible to simulate lower-dimensional Hamiltonians by means of atoms in strongly anisotropic external potentials \cite{RevModPhys.80.885,lewenstein-rmp-56-135}. For instance, in 2D, the algebraically correlated superfluid and the corresponding Berezinskii--Kosterlitz--Thouless phase transition have been observed for both bosons \cite{Hadzibabic2006,Clade2009,Tung2010,Plisson2011,Desbuquois2012,PhysRevLett.114.255302} and fermion pairs \cite{PhysRevLett.114.230401,PhysRevLett.115.010401}. A characteristic of these types of experiments is that the final setup might, due to an insufficient degree of anisotropy of the trap in comparison to the density of the system, still feature aspects of the 3D system. This anisotropy may be quantified by the relative spacing of energy levels for excitations related to different spatial directions, or by aspect ratios of trapping frequencies in different directions. For a finite anisotropy ratio, the system is, in general, in a dimensional crossover without a well-defined dimensionality. Whereas this is typically an unwanted effect in the quantum simulation of 2D and 1D Hamiltonians, it can also be seen as a possibility to implement new types of quantum matter with interesting properties. The aim of this paper is to investigate the few- and many-body properties of a particularly simple system, the Bose gas with pointlike interactions, in the dimensional crossover.

Consider for simplicity the crossover from 3D to 2D by means of compactifying the ``transverse'' $z$-direction. However, the following argumentation is not limited to this particular case.  As a particularly instructive transverse confinement, we delimit the $z$-direction by a potential well of length $L$. The boundary conditions at the end points may be chosen periodic, in which case we also say that the system is confined to a torus in $z$-direction. We may also restrict the system to a box potential with infinitely high walls given by
\begin{align}
 V_{\rm box}(z) = \begin{cases} 0 & 0\leq z \leq L\\ \infty & \text{else}\end{cases}.
\end{align}
In both cases, the trapping potential inside the well vanishes, and thus the confinement reduces to boundary conditions on the bosonic field $\phi(\tau,\vec{x})$, or, equivalently, the coordinates of the many-body wave function in a first-quantized formulation. These boundary conditions read
\begin{align}
 \label{m5}  \phi(\tau,x,y,z=0)=\phi(\tau,x,y,z=L)
\end{align}
for periodic boundary conditions (pbc) and
\begin{align}
 \label{m6} \phi(\tau,x,y,z=0)=\phi(\tau,x,y,z=L)=0
\end{align}
for the box. The boundary conditions lead to a quantization of energies and thus a discrete excitation spectrum in $z$-direction.

\begin{figure}[t]\centering
\includegraphics[width=\columnwidth]{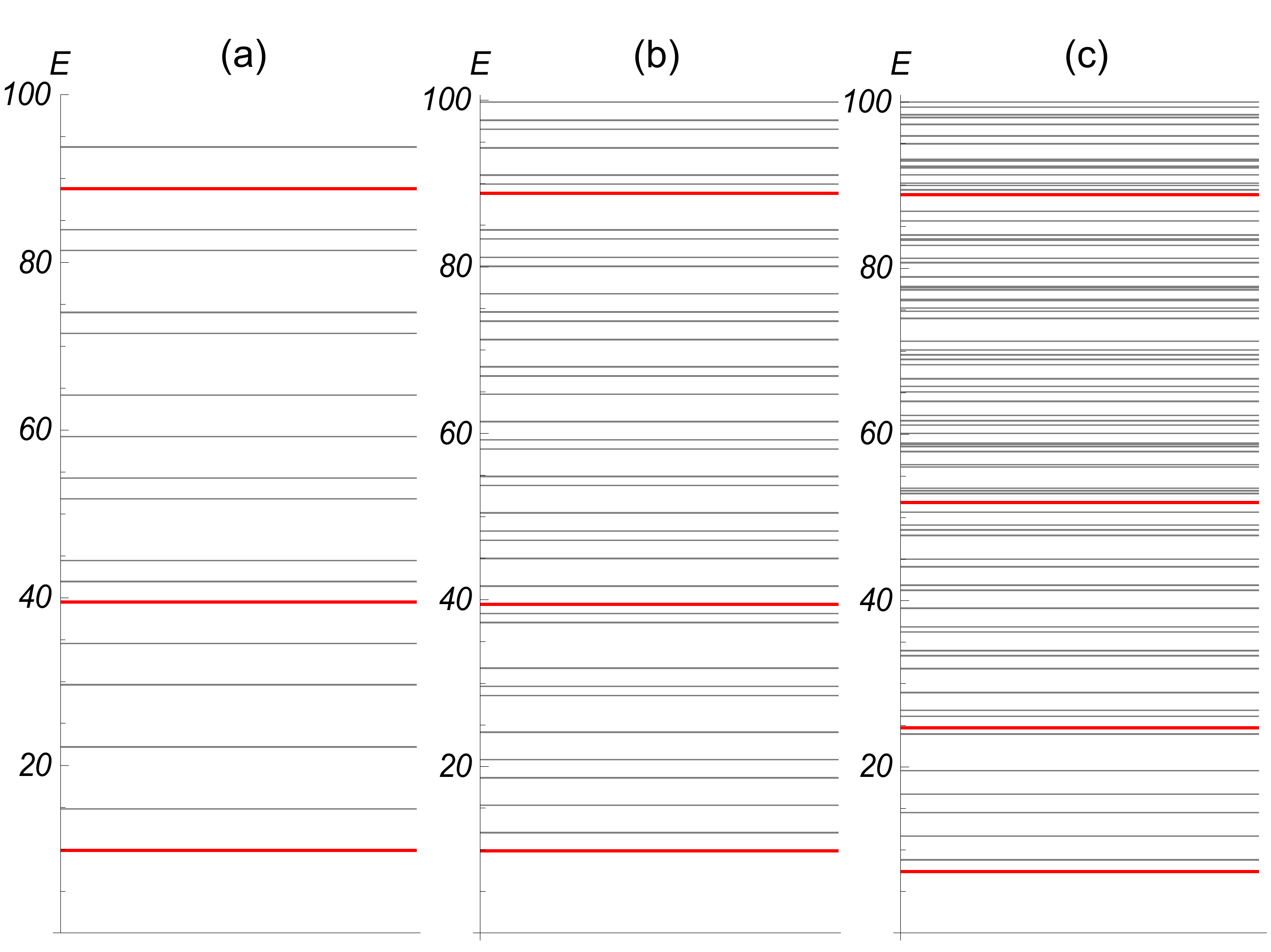}
\caption{Schematics of the energy spectrum in the dimensional crossover from 3D to 2D. The first two panels show the discrete eigenenergies for a noninteracting system in a 3D box potential which confines the system to the cuboid $(x,y,z)\in [0,L_{\rm x}]^2\times[0,L]$. In (a) we have set $L_{x}/L=2$, whereas (b) displays the case of $L_{x}/L=3$. As we increase the ratio $L_{x}/L$, an effective 2D continuum of states (thin gray lines) fills the region between the discrete spectrum of excitations in the transverse $z$-direction (thick red lines). The lowest excitations appear above the zero-point energy, which is generally not zero. In (c) we sketch the more realistic scenario of a system which is interacting and confined by a smooth trapping potential. Although the concrete mode spectrum and level spacing in (c) differ from (a) and (b), we find the same qualitative behavior. 
}
\label{fig:energy_spectrum}
\end{figure}

In a functional integral formulation of the many-body quantum system, the boundary conditions appear as restrictions on the appropriate function space. This can be related to energy levels of non-interacting excitations, and eigenstates of the interacting system are not needed for this purpose. In our setting the eigenfunctions in $z$-direction are superpositions of plane waves with momentum component $q_z$ and the boundary conditions (\ref{m5}) and (\ref{m6}) lead to a quantization of energies $E_z=\hbar^2q_z^2/2M$. For the torus we have $q_z\to k_n$ with
\begin{align}
 k_n = \frac{2\pi n}{L},\ n\in\mathbb{Z},
\end{align}
whereas in a box we have $E_z\to \hbar^2\kappa_n^2/(2M)$ with $\kappa_n = \frac{\pi n}{L},\ n=1,2,\dots$ Importantly, the box features a nonvanishing zero-point energy, $E_0$, whereas a trap with pbc does not. Since realistic confinements such as they appear in cold atom experiments are implemented by means of smoothly varying trapping potentials and not potential wells, our analysis can only reveal qualitative statements. For example, in case of a harmonic trap the function space consists of Hermite functions, again with a discrete energy spectrum. For an increasing anisotropy the level spacing in the $x$- and $y$-directions goes to zero as compared to the one in the $z$-direction. 
We visualize the previous statements in Fig. \ref{fig:energy_spectrum}.

In our setting we need to compare $L$ with the length scales associated to the many-body physics such as density and temperature. The regime where $L$ is much smaller than all those length scales is called the 2D limit, because the system will be in its ground state in the $z$-direction and the low-energy excitations are limited to the two-dimensional continuum $(q_x,q_y)$ in momentum space just above the zero-point energy $E_0$. The 3D system then macroscopically looks like a 2D system. In our analysis below we also address the question how the parameters of the effective 2D system are inherited from the 3D system.

Our formulation bears much resemblance to other systems where ``dimensional reduction'' occurs. This concerns, for example, the transition from effective quantum statistics to classical statistics as the temperature $T$ drops below the relevant energy scales of the system. In the Matsubara formalism, $1/T$ plays the same role as $L$ in our case. Another example is dimensional reduction of higher-dimensional Kaluza--Klein theories compactified on a torus. The main difference to systems in relativistic quantum field theory is the different non-relativistic dispersion relation in our setting.

This paper is organized as follows. In Sec. \ref{SecModel} we introduce the Bose gas in the dimensional crossover. In Sec. \ref{SecSuper} we study the superfluid phase transition in the crossover from 3D to 2D by means of the Functional Renormalization Group (FRG). The scattering properties in the 2D limit are studied with the FRG and T-matrix calculations in Sec. \ref{Sec2D}. In Sec. \ref{SecDiss} we discuss the relevance of our calculations for more realistic confinements and give an outlook on extensions of our approach. In Apps. \ref{AppTc2D} and \ref{AppTca2D} we study the behavior of the critical temperature in the 2D limit, and in Apps. \ref{AppFlow} and \ref{SecTmat} we derive the FRG flow equation in the dimensional crossover in detail and relate T-matrix, scattering vertex, and dimer propagator within the FRG framework.

\section{Model}\label{SecModel}
We study non-relativistic bosons with repulsive pointlike interactions. This provides an excellent description for ultracold quantum gases of Alkali atoms, where the precise form of interatomic interactions is irrelevant for the macroscopic physics and can thus be replaced by a pointlike s-wave coupling constant $\lambda_\Lambda$, which eventually relates to an experimentally measurable scattering length $a$. The model may, however, also be used as an effective description for non-relativistic bosonic degrees of freedom in other condensed matter setups.

We employ a functional integral formulation with the Euclidean microscopic action of the system given by
\begin{align}
\label{m1} S[\varphi]&=\int_X\Bigl[\varphi^*\Bigl(\partial_\tau-\frac{\nabla^2}{2M}-\mu\Bigr)\varphi+\frac{\lambda_\Lambda}{2}|\varphi|^4\Bigr].
\end{align}
Herein, the bosonic degrees of freedom are described by the complex scalar field $\varphi=\varphi(\tau,\vec{x})$. We abbreviate $\int_X=\int_0^\beta\mbox{d}\tau\int \mbox{d}^dx$, the number of spatial dimensions is $d$, and $\tau$ denotes imaginary time. The latter is compactified to a torus of circumference $\beta=(k_{\rm B}T)^{-1}$ with temperature $T$. The chemical potential and atomic mass are denoted by $\mu$ and $M$, respectively. The action is defined with respect to an ultraviolet cutoff $\Lambda$, which is on the order of the inverse van-der-Waals length for ultracold atoms. We use units such that $\hbar=k_{\rm B}=2M=1$.

The few- and many-body properties of the system described by the action (\ref{m1}) are captured by the partition function of the system. Equivalently, this information is encoded in the effective action or free energy $\Gamma[\phi;\mu,T,\lambda,\Lambda]$, which is a functional of the mean field $\phi(x)=\langle\varphi(x)\rangle$ in the presence of general inhomogeneous sources $J(x)$. In the following we employ the Functional Renormalization Group to deduce properties of $\Gamma$ within approximations. The approach is based on the exact evolution equation for the effective average action $\Gamma_k$ \cite{Wetterich:1989xg,Wetterich1993} given by
\begin{align}
\label{m2}\partial_k\Gamma_k=\frac{1}{2}\text{Tr}\left[(\Gamma_k^{(2)}+R_k)^{-1}\partial_k R_k\right],
\end{align}
where $k$ is a momentum scale such that $\Gamma_{k=\Lambda}=S$ for some ultraviolet scale $\Lambda$ and $\Gamma_{k=0}=\Gamma$ \cite{Berges:2000ew,Pawlowski20072831,Gies:2006wv,Schaefer:2006sr,Delamotte:2007pf,Kopietz2010,Metzner:2011cw,Braun:2011pp,Boettcher:2012cm}. The second functional derivative of $\Gamma$ is denoted by $\Gamma^{(2)}$, and $R_k$ is a regulator function, which can be chosen freely within some limitations \cite{Pawlowski:2015mlf}.

In order to approximate the right hand side of the flow equation (\ref{m2}) we apply the ansatz
\begin{align}
 \label{m3}\Gamma_k[\phi]=\int_X\Bigl(\phi^*(Z_k\partial_\tau-\nabla^2)\phi+U_k(|\phi|^2)\Bigr)
\end{align}
in the implementation of Refs. \cite{PhysRevB.77.064504,PhysRevA.77.053603,PhysRevA.79.013601}. Our main interest is in the effective potential $U_k$ that we evaluate here for a homogeneous field $\phi$. This is well suited for periodic boundary conditions, while for the box a field $\phi(z)\sim\sin(\pi z/L)$ should improve the approximation. (In principle, one could include different potentials for several of the low lying modes in the $z$-direction.) For the present analysis it is sufficient to employ a $\phi^4$-expansion for the effective average potential $U_k(\rho)$ according to 
\begin{align}
 \label{m4} U_k(\rho) = m^2_k \rho + \frac{\lambda_k}{2}\rho^2,
\end{align}
with U($1$)-invariant $\rho=\phi^*\phi$. In Eq. (\ref{m3}) we use a renormalized field $\phi$ chosen such that the prefactor of $\phi^*\nabla^2\phi$ is unity. This explains why the expectation value of $\phi$ can be nonzero even in the 2D-limit, where the Mermin--Wagner theorem \cite{Mermin1966, Hohenberg1967} forbids long-range order (LRO).

We denote the $k$-dependent minimum of $U_k(\rho)$ by $\rho_{0,k}$. The normal and superfluid phase of the system are distinguished by means of $\rho_{0,k\to0}$ being zero or nonzero, respectively. The ansatz in Eq. (\ref{m4}) has been successfully applied to describe the universal and non-universal features of the non-relativistic Bose gas in both 3D and 2D. In particular, the thermodynamic equation of state \cite{PhysRevA.79.063602,PhysRevA.85.063607}, critical exponents \cite{Tetradis1994541,Berges:2000ew}, and critical temperature $T_{\rm c}$ \cite{PhysRevA.79.013601,PhysRevA.85.063607} have been obtained in good quantitative agreement with other theoretical approaches \cite{Prokofev2001,PhysRevA.66.043608,Holzmann2007,Holzmann2008} and with experiment \cite{Hung2011,PhysRevLett.107.130401,Zhang1070}. Extension of the ansatz (\ref{m4}) for $\Gamma_k$ of the 2D Bose gas have been devised in Refs. \cite{PhysRevB.64.054513,PhysRevLett.102.190401,PhysRevE.90.062105} to capture, for instance, the emergence of Popov's hydrodynamic description at low energies and the essential scaling at the BKT transition.

Given the success of the ansatz (\ref{m4}) to provide a qualitative and mostly quantitative correct description of the 3D and 2D non-relativistic Bose gas, it is natural to apply it to the question of the dimensional crossover from 3D to 2D. For spatial dimensions $2<d\leq 3$, the system features LRO below the critical temperature $T_{\rm c}$. For $d=2$, a superfluid phase with power-law decay of correlations and nonzero superfluid density $\rho_{0,k\to0}$ exists below a critical temperature $T_{\rm c}$ \cite{Berezinskii1971,Berezinskii1972,Kosterlitz1973,Kosterlitz1974}. Investigating the location of the minimum $\rho_{0,k}$ of the effective average potential in the limit $k\to0$ thus allows for a unified description of the superfluid transition for all $2\leq d \leq3$.

The derivative expansion employed in Eq. (\ref{m3}) is spatially isotropic in the sense that excitations in the planar and transverse directions are treated equally. More generally, the kinetic energy of excitations may be approximated as
\begin{align}
 E_{\rm kin}(\vec{q},q_z) = \vec{q}^2 + \xi q_z^2,
\end{align}
where $\vec{q}$ is the momentum vector in the plane, and $q_z$ is the component in the transverse (confined) direction. We call this more general ansatz the anisotropic derivative expansion. In App. \ref{AppFlow} we show that it reproduces the same results as the ansatz in Eq. (\ref{m3}) with $\xi=1$, because $\xi$ remains of order unity and is not exceptionally small. Accordingly, excitations in the transverse direction with momentum $q_z\sim L^{-1}$ are still energetically costly for $L\to 0$ and thus decouple from the low-energy physics. As soon as they are decoupled, however, the precise value of the prefactor $\xi$ is unimportant. The detailed discussion is presented in App. \ref{AppFlow}.

It is instructive to study the role of $\rho_0$ in the dimensional crossover in terms of the effective action. For this consider the 3D effective action ansatz
\begin{align}
 \nonumber\Gamma[\phi] = \int_X \int_0^L\mbox{d}z\Bigl(&\phi^*(Z_k\partial_\tau-\nabla-\partial_z^2+m^2)\phi\\
&+\frac{\lambda_{\rm 3D}}{2}|\phi|^4\Bigr),
\end{align}
where we denote $X=(\tau,\vec{x})$ with $\vec{x}=(x,y)$. The field $\rho=|\phi|^2$ has the physical dimension of a 3D number density. The ground state $\phi_0(\tau,\vec{x},z)$ of the 3D system with $L= \infty$ is given by a constant value $\phi_0$ with arbitrary complex phase. The same holds for pbc, where we can write
\begin{align}
 \phi_0^{(\rm pbc)}(\tau,\vec{x},z) = \chi_0 \frac{1}{\sqrt{L}},
\end{align}
with $|\chi_0|^2$ having the dimension of a 2D number density. The effective action evaluated for this configuration reads
\begin{align}
  \Gamma[\phi_0^{(\rm qbc)}]= \int_X \Bigl( m^2|\chi_0|^2+ \frac{(\lambda_{\rm 3D}/L)}{2}|\chi_0|^4\Bigr).
\end{align}
It resembles the one of a 2D Bose gas with coupling constant $\lambda_{\rm 2D}=\lambda_{\rm 3D}/L$. (Since $\lambda_{\rm 3D}$ has dimension of a length, $\lambda_{\rm 2D}$ is dimensionless.) For a confinement in a box we have an inhomogeneous ground state given by
\begin{align}
\phi_{0}^{(\rm box)}(\tau,z,\vec{x}) = \chi_0 \sqrt{\frac{2}{L}}\sin\Bigl(\frac{\pi z}{L}\Bigr).
\end{align}
For this configuration we have
\begin{align}
  \label{m23} \Gamma[\phi_0^{(\rm box)}]  &= \int_X \Bigl(\Bigl(m^2+\frac{\pi^2}{L^2}\Bigr)|\chi_0|^2 + \frac{\lambda_{\rm 2D}}{2}|\chi_0|^4\Bigr)
\end{align}
with $\lambda_{\rm 2D}=I_2\frac{\lambda_{\rm 3D}}{L}$, $I_2=\frac{4}{L} \int_0^L \mbox{d}z \sin^4(\frac{\pi z}{L})=\frac{3}{2}$. Hence, for the confinement with sharp boundary conditions we find a nontrivial prefactor in the translation between the coupling constants on the mean-field level, namely $\lambda_{\rm 2D} = 3\lambda_{\rm 3D}/(2L)$. The coherence properties of ultracold Bose gases in the dimensional crossover with confining box potentials have been discussed in Ref. \cite{PhysRevA.68.043603}.

The FRG approach has been applied to study the effect of compactified dimensions on many-body system both in several relativistic and non-relativistic setups. Finite volume effects, i.e., when \emph{all} spatial directions are confined to a cube of size $[0,L]^d$, have been studied for quark-meson-models \cite{2009EPJC...63..443B,2011EPJC...71.1576B,2012PhLB..713..216B,2014PhRvD..90e4012T} and the 3D BCS-BEC crossover \cite{2011PhRvA..84f3616B}. In particular, Ref. \cite{2014PhRvD..90e4012T} highlights the difference in periodic and antiperiodic boundary conditions. The latter result in a nonvanishing zero-point energy similar to the confinement in a box potential discussed above. In Section \ref{SecDiss} we discuss the role of the zero-point energy played in the FRG setup on the basis of our findings for the dimensional crossover.

\section{Superfluid transition}\label{SecSuper}
The flow of the effective average action $\Gamma_k$ is initialized at some ultraviolet momentum scale $k=\Lambda$, where $\Gamma_\Lambda$ coincides with the microscopic action (\ref{m1}) of a 3D Bose gas. The compactification in $z$-direction introduces a new scale to the 3D system, which is defined by the size $L$ of the potential well. In particular, note that if $L^{-1}$ is much larger than the many-body scales, the system is effectively two-dimensional. We examine this scenario in Sec. \ref{Sec2D}. However, the UV scale $\Lambda$ is always chosen in a way that $\Lambda\gg (L^{-1},\mu^{1/2},T^{1/2})$, i.e., we start with a 3D system with 3D coupling constants. 
The initial value for the effective potential is given by
\begin{align}
 U_\Lambda(\rho) = \frac{\lambda_{\Lambda}}{2}(\rho - \rho_{0,\Lambda})^2
\end{align}
with $\rho_{0,\Lambda}=\mu/\lambda_\Lambda$. In practice, we initialize the flow at the UV scale $\Lambda/\sqrt{\mu}=10^3$, and stop at the final scale $k_{\rm f}=\Lambda e^{-10}$, which is much smaller than the many body scales.

The flow equation of the effective average potential $U_k$ factorizes into a universal $L$-independent part $G_1(T)$ and a crossover function $F(\tilde{L})$ according to
\begin{align}
\label{st1a}
k\frac{\partial}{\partial k}U_k(\rho)&=\frac{k^{5}}{4\pi Z_k}G_1(T)F(\tilde{L})+\eta \rho U_k'(\rho).
\end{align}
The universal part reads
\begin{align}
 \nonumber G_1(T)&=\Bigl(\sqrt{\frac{1+w_1}{1+w_2}}+\sqrt{\frac{1+w_2}{1+w_1}}\Bigr)\\
\label{st1b} &\times \Biggl(\frac{1}{2}+N_B\Bigl(\frac{k^2\sqrt{(1+w_1)(1+w_2)}}{Z_k}\Bigr)\Biggr),
\end{align}
with $w_1=U'/k^2$, $w_2=(U'+2\rho U'')/k^2$, and the Bose function $N_B(x)=(e^{x/T}-1)^{-1}$. Here a prime denotes a derivative with respect to $\rho$. Further, $\eta$ constitutes the $k$-dependent anomalous dimension associated with field rescaling. For a detailed derivation of Eq. (\ref{st1a}) see Appendix \ref{AppFlow}.

\begin{figure}[t]\centering
\includegraphics[width=8cm]{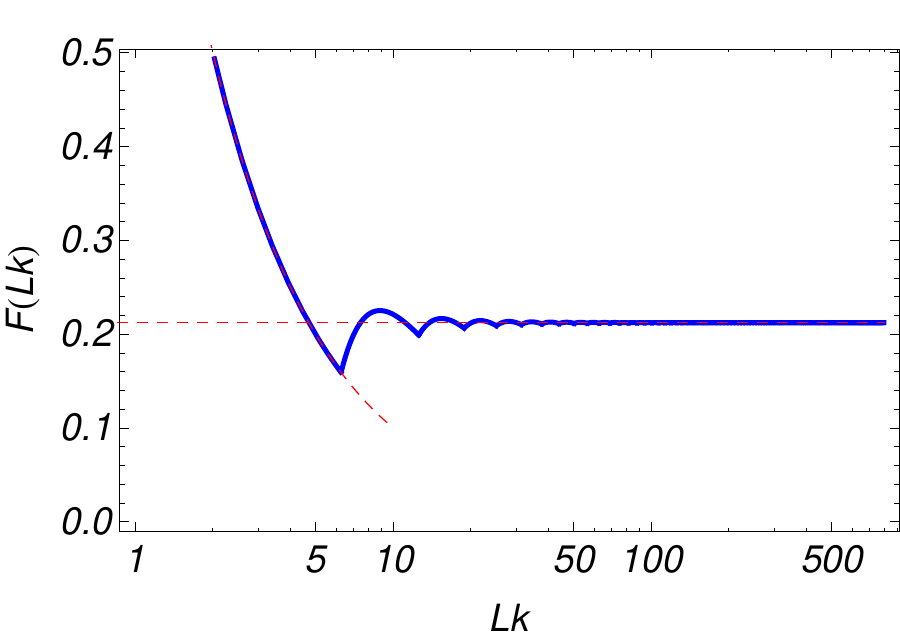}
\caption{The crossover function $F(\tilde{L})$ modifies the flow equations of the 3D system. For small $\tilde{L}=Lk$ it diverges like $\tilde{L}^{-1}$ and the 2D flow equations are recovered. For large $\tilde{L}$ it approaches a constant and the system behaves three-dimensional. In the FRG approach we start at an ultraviolet scale $k_\Lambda=\Lambda$ with $L\Lambda\gg 1$, so that the description of the microscopic theory is three-dimensional.}
\label{fig:crossover_function}
\end{figure}

The influence of the trap is encoded in the crossover function $F(\tilde{L})$, which for pbc is given by
\begin{align}
\label{st1}
\nonumber F_{\rm pbc}(\tilde{L})&=\frac{2N+1}{\tilde{L}}\left[1-\frac{\eta}{4}-\frac{1}{\tilde{L}^2}\left(1-\frac{\eta}{2}\right)\frac{4\pi^2}{3}N(N+1)\right.\\
&\left.-\frac{\eta}{\tilde{L}^4}\frac{4\pi^4}{15}N(N+1)(-1+3N+3N^2)\right],
\end{align}
with $\tilde{L}=Lk$, $N=\lfloor \frac{\tilde{L}}{2\pi}\rfloor$, and $\lfloor x\rfloor$ being the largest integer $<x$. The function $F$ interpolates between the 3D and 2D flow equations via the two limiting cases
\begin{align}
 \label{st2} F(\tilde{L}) = \begin{cases} \frac{2}{3\pi}\left(1-\frac{\eta}{5}\right) & (\tilde{L}\gg 1) \\\tilde{L}^{-1}\left(1-\frac{\eta}{4}\right) & (\tilde{L} \ll 1)\end{cases}.
\end{align}
We plot the crossover function for pbc in Fig. \ref{fig:crossover_function}. One observes a rather sharp transition from the 3D behavior for $k>\frac{2\pi}{L}$ to the 2D behavior for $k<\frac{2\pi}{L}$. The oscillatory behavior in the transition region is rapidly damped. The flow equation for the dashed lines corresponds precisely to the flow in 3D or 2D with the same truncation. Dimensional reduction is very effectively realized by the flow with a reasonable approximation simply switching from 3D to 2D as $k$ decreases below $2\pi/L$. A similar sharp transition has been observed in the $\tau$-direction for the transition between quantum and classical statistics \cite{Tetradis:1992xd,Tetradis:1993bx}.

From Eq. (\ref{st1a}), the flow equations of $U'_k(\rho_0)$ and $\lambda_k=U''_k(\rho_0)$ follow by simple differentiation with respect to $\rho$. The flow of the location of the minimum $\rho_{0,k}$ is obtained by
\begin{align}
  \frac{\partial}{\partial k}\rho_{0,k}=-\lambda_k^{-1}\frac{\partial}{\partial k}U'_k\bigg|_{\rho_0}.
\end{align}
By following the flow of $\rho_{0,k}$ for $k\rightarrow 0$ we deduce whether $\rho_{0,0}>0$ (superfluid phase) or $\rho_{0,0}=0$ (disordered phase). This allows us to extract the superfluid transition temperature $T_c$ for all trap sizes $L$. We determine the density by a suitable flow equation, see Ref. \cite{PhysRevA.77.053603}.

\begin{figure}[t]\centering
\includegraphics[width=8cm]{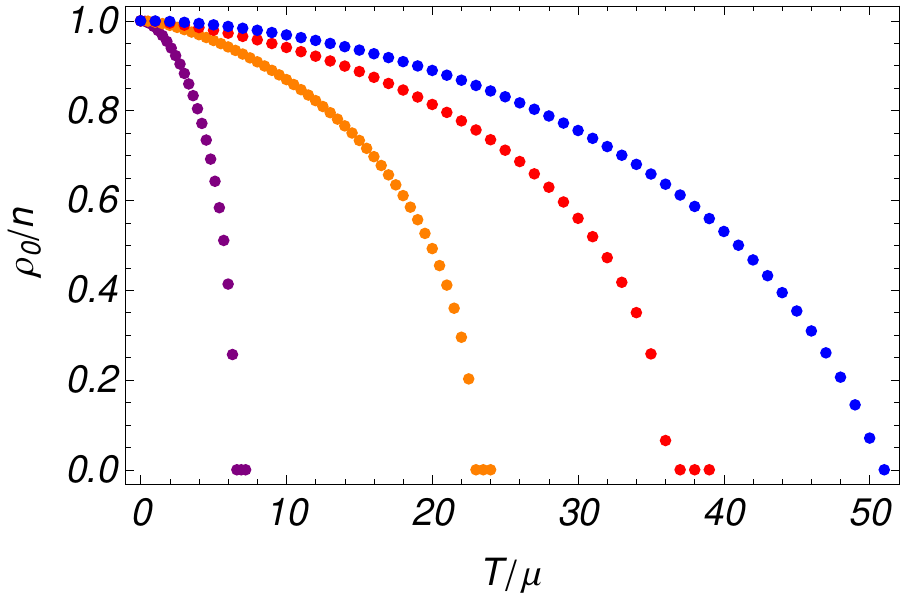}
\caption{Superfluid fraction $\rho_0/n$ at $k\to 0$ for different values of $L$. Larger values of $L$ correspond to the curves further to the right. We directly see that the critical temperature, here in units of the chemical potential $\mu$, is higher in more 3D-like systems. For all compactification lengths $L$ we recover the correct limit $\rho_0/n|_{T\to0}=1$, where the whole gas is superfluid due to Galilean invariance. The curves correspond to a coupling strength $g_{\rm 3D}\sqrt{\mu}=0.025$.}
\label{fig:rho_vs_T}
\end{figure}

We plot the superfluid fraction $\rho_0/n$ for different values of $L$ in Fig. \ref{fig:rho_vs_T}. Note that even for a 2D system the ratio $\rho_0/n$ of the 3D quantities $\rho_0$ and $n$ gives the superfluid fraction, since $\rho_0^{\rm 2D}=L\rho_0^{\rm 3D}|_{L\to 0}$ and $n^{\rm 2D}= Ln^{\rm 3D}|_{L\to 0}$, so that $L$ drops out. For vanishing temperature the whole gas becomes superfluid irrespective of $L$, and develops a small condensate depletion due to interactions \cite{PhysRevA.77.053603}.

\begin{figure}[t]\centering
\includegraphics[width=8cm]{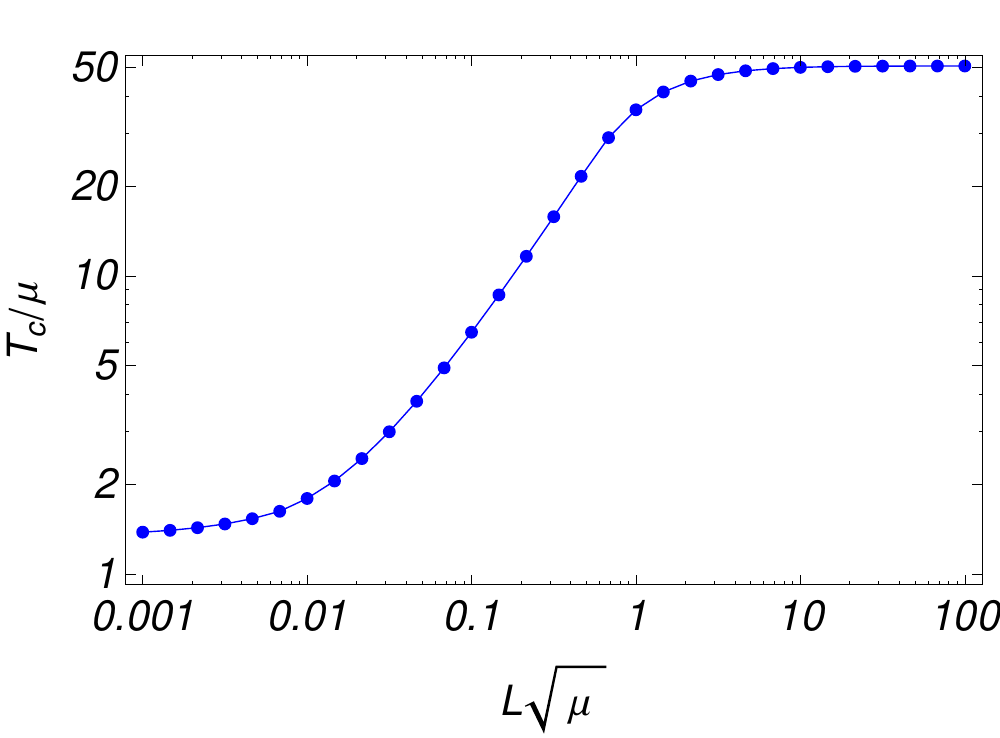}
\caption{Dimensional crossover of the superfluid critical temperature $T_{\rm c}$  from 2D to 3D. For large $L$ the superfluid transition goes along with Bose--Einstein condensation. The transition temperature is substantially lower in a smaller system, i.e., when $L$ is smaller. For the crossover we fix the 3D coupling constant to the same value as in Fig. \ref{fig:rho_vs_T}. In experiments with ultracold atoms, $g_{\rm 3D}$ is usually controlled by a magnetic Feshbach resonance. 
}
\label{fig:Tc_vs_L}
\end{figure}

In Fig. \ref{fig:Tc_vs_L} we show the superfluid transition temperature $T_{\rm c}$ for the whole 2D--3D dimensional crossover. For large $L$, $T_{\rm c}$ approaches the Bose--Einstein condensation (BEC) temperature $T_{\rm BEC}$ of the interacting gas given by
\begin{align}
\nonumber T_{\rm BEC}&=T_{\rm c,{\rm id}}+\Delta T_{\rm c},\\
T_{\rm c,{\rm id}}&=\frac{2\pi\hbar^2}{Mk_{\rm B}}\Bigl(\frac{n}{\zeta(3/2)}\Bigr)^{2/3}=6.625n^{2/3}.
\end{align}
The interaction induced shift of the critical temperature \cite{PhysRevLett.79.3549,PhysRevLett.83.1703} for sufficiently small gas parameter $a_{\rm 3D}n^{1/3}$ reads
\begin{align}
 \label{stcrit2}\frac{\Delta T_{\rm c}}{T_{\rm c,{\rm id}}}=\kappa\ a_{\rm 3D}n^{1/3}.
\end{align}
Within our calculation we have $\kappa=2.1$ \cite{PhysRevA.77.053603}, which compares well with the Monte Carlo result \cite{PhysRevLett.83.2687,PhysRevLett.87.120401,PhysRevLett.87.120402}. In our setup we define $g_{\rm 3D}$ by
\begin{align}
 \label{stcrit2b} g_{\rm 3D} = 8\pi a_{\rm 3D} = \lambda_{{\rm 3D,} k\to 0}.
\end{align}
In terms of the chemical potential we find $T_c/\mu=51$ for $g_{\rm 3D}\sqrt{\mu}=0.025$, which corresponds to $\mu_{\rm c}=1.9g_{\rm 3D}n$ at the transition. Note that Hartree--Fock theory predicts $\mu=2g_{\rm 3D}n$ above the transition temperature.

As $L$ is lowered towards the 2D case, the critical temperature decreases and is reduced by a substantial factor as one goes through the crossover. Here we perform the dimensional crossover for a fixed value of $a_{\rm 3D}=g_{\rm 3D}/(8\pi)$, since this quantity is typically fixed by the magnetic field in cold atom experiments. This, in turn, fixes the initial 3D coupling $\lambda_{\Lambda}^{\rm 3D}$ through Eq. (\ref{2d12}), and leads to a characteristic behavior of $T_{\rm c}$ in the 2D limit of small $L$. For this note that for $L\gtrsim \Lambda^{-1}$ the flow in Eq. (\ref{st1a}) is solely determined by $F(\tilde{L})\simeq \tilde{L}^{-1}(1-\frac{\eta}{4})$, and thus resembles a truly 2D RG flow. In particular, this applies to the initial value of the coupling which for $L\to 0$ is given by
\begin{align}
\label{abcd} \lambda_{{\rm 2D},\Lambda}\Bigr|_{L\Lambda\gtrsim 1} =\frac{1}{L} \lambda_{{\rm 3D},\Lambda} \approx 8\pi \frac{a_{\rm 3D}}{L},
\end{align}
see Eq. (\ref{8pi}). Hence, as we decrease $L$ in this regime, we simulate a whole class of strongly-interacting 2D Bose gases, each corresponding to a certain coupling $\lambda_{\rm 2D}(L)$ and critical temperature $T_{\rm c}(L)$. Consequently, $T_{\rm c}$ does not fully saturate for small $L$, as can be seen in Fig. \ref{fig:Tc_vs_L}, although the dependence becomes weak.

In cold atom experiments, the 2D coupling strength is commonly expressed by the coupling constant $\tilde{g}=(M/\hbar^2)g_{\rm 2D}=g_{\rm 2D}/2$. Due to the logarithmic energy dependence of the T-matrix in 2D, the coupling $g_{\rm 2D}$ needs to be defined with respect to a fixed energy or momentum scale (in contrast to $a_{\rm 2D}$). On the other hand, as this scale dependence is only logarithmic, $\tilde{g}$ appears to be effectively constant in experiments. One convenient choice consists in defining the coupling at the momentum scale $k_n=\sqrt{n_{\rm 2D}}$ related to the 2D density $n_{\rm 2D}=L n_{\rm 3D}$. For studying the phase transition, $T> \mu$ sets the appropriate scale and thus for a true 2D system we define
\begin{align}
 \label{ab} g_{\rm 2D} = 2 \tilde{g} = \lambda_{{\rm 2D}, k=\sqrt{T}}
\end{align}
in the FRG approach. Another convenient choice consists in identifying the expression in Eq. (\ref{abcd}) with $g_{\rm 2D}$.

For a true 2D system with small $\tilde{g}$, the critical phase space density for the BKT transition is $n\lambda_T^2=\log(\xi/\tilde{g})$, where $\lambda_T^2=\frac{2\pi\hbar^2}{Mk_BT}=\frac{4\pi}{T}$. Monte Carlo calculations yield $\xi=380$ \cite{Prokofev2001,PhysRevA.66.043608}.
Hence the 2D critical temperature reads
\begin{align}
 \label{stcrit3}T_{\rm BKT}=\frac{4\pi n_{\rm 2D}}{\log(\xi/\tilde{g})}.
\end{align}
By employing $\mu_c/T=(\tilde{g}/\pi)\log(\xi_\mu/\tilde{g})$ for the critical chemical potential, we may also express the critical temperature for small $\tilde{g}$ by
\begin{align}
 \label{stcrit4}\frac{T_{\rm BKT}}{\mu}= \frac{\pi}{\tilde{g}}\frac{A}{\log (\xi_\mu/\tilde{g})}.
\end{align}
Here $A=1$ and $\xi_\mu=13.2$ from Monte Carlo computations \cite{Prokofev2001}. In our setup, the values of $A$ and $\xi_\mu$ mildly depend on the value of $k_{\rm f}$ introduced in the beginning of this section, see the discussion in Ref. \cite{PhysRevA.79.013601} and Appendix \ref{AppTc2D}. For the choice of $k_{\rm f}$ applied here we find $A\simeq1$ and $\xi_\mu\simeq6$. The parametric form in Eq. (\ref{stcrit4}) has also been obtained with the FRG in Ref. \cite{PhysRevA.85.063607}, where $T_{\rm BKT}$ was determined in a different manner than employed here, yielding $A=0.982$ and $\xi_\mu=9.48$.

For experiments with ultracold 2D gases it might be relevant that a finite extension in the $z$-direction leads to an enhanced superfluid transition temperature. Depending on the system this influence can be quite substantial. For a related study in the context of BCS-superfluidity of two-component fermions see Ref. \cite{PhysRevB.90.214503}, where an enhancement of the critical temperature in a quasi-2D geometry has been found. Furthermore, since in 3D we also have a macroscopic occupation of the lowest energy state, that is a condensate, there might still be some residual condensate for small yet finite $L$ and nonvanishing temperatures.

To investigate the question whether the partial influence of the third dimension influences the critical temperature of a (supposed to be) 2D system, we perform the dimensional crossover for a fixed value of $a_{\rm 2D}$. Note that this is in contrast to the data shown in Figs. \ref{fig:rho_vs_T} and \ref{fig:Tc_vs_L}, where $a_{\rm 3D}$ is kept fixed. In App. \ref{AppTc2D} we provide a discussion of the dimensional crossover of $T_{\rm c}/\mu$ for small $L$ and fixed $a_{\rm 2D}$. We find that for sufficiently small $L/a_{\rm 2D}$, the critical temperature coincides with that of the true 2D system, and is then \emph{enhanced} for larger values of $L$.

\section{2D limit}\label{Sec2D}

\subsection{Effective 2D coupling constant}

We have seen that the flow of couplings can be described
by a 2D flow whenever the scale $k$ is below
an effective cutoff $\Lambda_{\rm eff}\ll L^{-1}$. 
In particular, if $L^{-1}$ is much larger than the many-body scales $\sqrt{\mu}$ and $\sqrt{T}$, the many-body 
system can be described by an effective 2D model with
UV-cutoff $\Lambda_{\rm eff}$. We call this scenario the 2D limit. The aim of this section is to relate $\lambda_{\rm 2D}(\Lambda_{\rm eff})$ of the effective 2D action to the
3D scattering length. By integrating the flow equations between $\Lambda$ and $\Lambda_{\rm eff}$ in vacuum, we obtain the relation between $a_{\rm 2D}$ and $a_{\rm 3D}$, which depends on the  particular compactification or trap potential. Once the renormalized coupling $\lambda_{\rm 2D}(\Lambda_{\rm eff})$ is specified, it allows to universally determine the many-body properties of the effective 2D system, and no further knowledge of the 3D system is required.

In terms of the flow equation (\ref{st1a}), the 2D limit manifests itself in the fact that the prefactor $F(\tilde{L})$ approaches the regime $\tilde{L}=Lk\ll 1$ much earlier than any of the many-body scales is resolved. We introduce the momentum scale $\Lambda_{\rm eff}$ via
\begin{align}
\label{2d2} \Lambda \gg L^{-1} \gg \Lambda_{\rm eff} \gg \sqrt{\mu},\sqrt{T}.
\end{align}
The scale $\Lambda_{\rm eff}$ is arbitrary within these constraints and serves as an UV-cutoff of the effective 2D theory.

In the following we consider the regime  $k\in[\Lambda_{\rm eff},\Lambda]$ of the RG flow. It is characterized by $Z_k=1$, $\eta=\rho_{0,k}=N_B=0$, leading to a simple form of Eq. (\ref{st1a}), which reads
\begin{align}
 \label{2d3} \dot{U}(\rho) = \frac{k^5}{8\pi} F_0(\tilde{L}) \Bigl(\sqrt{\frac{1+w_1}{1+w_2}}+\sqrt{\frac{1+w_2}{1+w_1}}\Bigr),
\end{align}
where the dot denotes a derivative with respect to $t=\log(k/\Lambda)$. The crossover function $F|_{\eta=0}=F_0$ is evaluated for $\eta=0$, see Eq. (\ref{fe38}) for pbc. In particular, for $\tilde{L}\ll1$ we are left with
\begin{align}
 \label{2d4} \dot{U}(\rho) = \frac{1}{L} \frac{k^4}{8\pi}  \Bigl(\sqrt{\frac{1+w_1}{1+w_2}}+\sqrt{\frac{1+w_2}{1+w_1}}\Bigr) = \frac{1}{L}\dot{U}_{\rm 2D}.
\end{align}
This is, up to the prefactor $L^{-1}$, the flow equation of the 2D potential defined via $U_{\rm 3D}=\frac{1}{L} U_{\rm 2D}$ and $\rho_{\rm 3D}=\frac{1}{L}\rho_{\rm 2D}$. Note that with this definition the quantities $w_1 = U_{\rm 3D}'/k^2 =  U_{\rm 2D}'/k^2$ and $w_2=(U_{\rm 3D}' +2 \rho_{\rm 3D} U_{\rm 3D}'')/k^2=(  U_{\rm 2D}'+2 \rho_{\rm 2D} U_{\rm 2D}'')/k^2$ are invariant. We define $\lambda^{\rm 2D}=U_{\rm 2D}''(0)$ in vacuum such that the effective 2D coupling strength at scale $\Lambda_{\rm eff}$ is given by
\begin{align}
 \label{2d6} \lambda_{\Lambda_{\rm eff}}^{\rm 2D} = \frac{1}{L} \lambda_{\rm \Lambda_{\rm eff}}^{\rm 3D}.
\end{align}

The running of the 3D coupling constant $\lambda_k=\lambda^{\rm 3D}_k=U_{\rm 3D}''(0)$ in vacuum is found to be
\begin{align}
\label{2d5} \dot{\lambda}_k = \frac{k^5}{4\pi} F_0(\tilde{L}) \frac{\lambda_k^2}{k^4}.
\end{align} 
In order to solve this equation we write
\begin{align}
 \label{2d9}  \partial_k \frac{1}{\lambda_k} = -\frac{1}{4\pi}F_0(\tilde{L})
\end{align}
and arrive at
\begin{align}
 \label{2d10b} \frac{1}{\lambda_{\Lambda_{\rm eff}}^{\rm 3D}}-\frac{1}{\lambda_{\Lambda}^{\rm 3D}}=\frac{1}{4\pi L}\int_{L\Lambda_{\rm eff}}^{L\Lambda}{\rm d}\tilde{L}\ F_0(\tilde{L}).
\end{align}
Using Eq. (\ref{2d6}) we obtain
\begin{align}
 \label{2d11}\frac{1}{\lambda_{\Lambda_{\rm eff}}^{\rm 2D}}=\frac{L}{\lambda_\Lambda^{\rm 3D}}+\frac{1}{4\pi}\int_{L\Lambda_{\rm eff}}^{L\Lambda}{\rm d}\tilde{L}\ F_0(\tilde{L}).
\end{align}

The coupling constants $\lambda_\Lambda^{\rm 3D}$ and $\lambda_{\Lambda_{\rm eff}}^{\rm 2D}$ can be related to the 3D and 2D scattering lengths $a_{\rm 3D}$ and $a_{\rm 2D}$ by means of the formulas
\begin{align}
 \label{2d12} \frac{1}{\lambda_{\Lambda}^{\rm 3D}} &= -\frac{\Lambda}{6\pi^2} + \frac{1}{8\pi a_{\rm 3D}},\\
 \label{2d13} \frac{1}{\lambda_{\Lambda_{\rm eff}}^{\rm 2D}} &= -\frac{1}{8\pi}\log(\Lambda^2_{\rm eff} a_{\rm 2D}^2) +\frac{1}{8\pi},
\end{align}
see Eqs. (\ref{t3a}) and (\ref{t3b}).
Inserting these relations into Eq. (\ref{2d11}) we find
\begin{align}
a_{\rm 2D} = L \exp\Bigl\{-\frac{1}{2}\frac{L}{a_{\rm 3D}}+\Phi\Bigr\}
\end{align}
with
\begin{align}
 \Phi = \frac{2L\Lambda}{3\pi}-\log(L\Lambda_{\rm eff})+\frac{1}{2}-\int_{L\Lambda_{\rm eff}}^{L\Lambda}\mbox{d}\tilde{L}\ F_0(\tilde{L}).
\end{align}
Note that the artificial scales $\Lambda_{\rm eff}$ and $\Lambda$ drop out due to the particular limits of $F(\tilde{L})$ in Eq. (\ref{st2}) and we can deduce $a_{\rm 2D}$ purely from the values of $a_{\rm 3D}$ and $L$. This is true for any crossover function $F(\tilde{L})$ that satisfies (\ref{st2}) and approaches these limits sufficiently fast in the variable $\tilde{L}$.

For a confinement along the z-direction with pbc we numerically find $\Phi=0$ and are thus left with
\begin{align}
 \label{2d23} a_{\rm 2D}^{\rm (pbc)} = L\ \exp\Bigl\{-\frac{1}{2}\frac{L}{a_{\rm 3D}}\Bigr\}.
\end{align}
Since $\Lambda$ dropped out of the formula, the new effective cutoff of the 2D theory is given by $L^{-1}$. In particular, this formula should be compared with the result in a harmonic trap \cite{PhysRevA.64.012706,RevModPhys.80.885,Levinsen2015} given by
\begin{align}
 \label{2d24} a_{\rm 2D} = \ell_z \sqrt{\frac{\pi}{A}}\exp\Bigl\{-\sqrt{\frac{\pi}{2}}\frac{\ell_z}{a_{\rm 3D}}\Bigr\}
\end{align}
with $A=0.905$ and $\ell_z=\sqrt{\hbar/M \omega_0}$ the oscillator length. We may identify the effective length scale $\ell_z^{\rm eff}=L/\sqrt{2\pi}$ and an effective constant $A_{\rm eff}$ to match our results to Eq. (\ref{2d24}) according to
\begin{align}
 \label{2d25} a_{\rm 2D} = \ell_z^{\rm eff} \sqrt{\frac{\pi}{A_{\rm eff}}}\exp\Bigl\{-\sqrt{\frac{\pi}{2}}\frac{\ell_z^{\rm eff}}{a_{\rm 3D}}\Bigr\}.
\end{align}
We have $A_{\rm eff}^{(\rm pbc)}=\frac{1}{2}$ for the case of pbc.

Note also that Eqs. (\ref{2d6}) and (\ref{2d12}) imply
\begin{align}
 \label{8pi} \lambda_{{\rm 2D},\Lambda_{\rm eff}}= \frac{1}{L} \lambda_{{\rm 3D},\Lambda_{\rm eff}} = \frac{1}{\frac{L}{8\pi a_{3D}}-\frac{L\Lambda_{\rm eff}}{6\pi^2}}.
\end{align}
If $a_{\rm 3D}/L\to 0$ is sufficiently small, we can ignore the term of order $\mathcal{O}(L\Lambda_{\rm eff})$ in the denominator. Furthermore, neglecting the logarithmic running of the coupling, we can identify $g_{\rm 2D}\approx \lambda_{{\rm 2D},\Lambda_{\rm eff}}$ and thus arrive at
\begin{align}
 \label{2d25b} g_{\rm 2D} \approx 8\pi \frac{a_{\rm 3D}}{L}.
\end{align}
Written in terms of $\tilde{g}=g_{\rm 2D}/2$ and $\ell_z^{\rm eff}$ this assumes the familiar form
\begin{align}
 \label{2d26} \tilde{g} \approx \sqrt{8\pi} \frac{a_{\rm 3D}}{\ell_z^{\rm eff}}
\end{align}
known from the case of harmonic confinement \cite{RevModPhys.80.885}.

\subsection{T-matrix}
The scattering properties of bosons in the dimensional crossover are naturally included in the FRG flow equations in the limit of vanishing density and temperature. Here we compare and benchmark our findings with the quantum mechanical T-matrix approach to low-energy scattering, where the particle-particle loop is integrated directly without the use of a flow equation. In App. \ref{SecTmat} we give a detailed discussion of the interrelation of both approaches, and limit this presentation to the main results. The fact that the flow equation for the T-matrix in vacuum can be solved exactly is related to a decoupling property of the two-body sector in this particular system; see Ref. \cite{Floerchinger:2013tp} for an extensive analysis. The route to determining the scattering properties of two particles in a box potential via the T-matrix has also been explored in Ref. \cite{2015JPhB...48b5302Y} and Eqs. (\ref{2d23}) and (\ref{fff}) agree with the results therein for $L=2\pi R$.

For computing the T-matrix we utilize a sharp momentum cutoff which equips the momentum integration with a UV cutoff scale $\Lambda$. This regularization differs from the use of the regulator $R_k(Q)$ in Eq. (\ref{fe4}). We make this explicit by a subscript (sh) to the microscopic coupling, $\lambda_{\Lambda}^{(\rm sh)}$. One result of the analysis in App. \ref{SecTmat} is to relate $\lambda_\Lambda^{(\rm sh)}$ to $\lambda_\Lambda$, the latter being used in the FRG analysis in Eqs. (\ref{2d12}) and (\ref{2d13}). Note at this point that it is difficult to obtain the additional term $+\frac{1}{8\pi}$ in Eq. (\ref{2d13}) without computing the energy-dependence of the T-matrix.

The low-energy T-matrix $T(E)$ in $d$ noncompact dimensions is defined as
\begin{align}
 \label{tmat} \frac{1}{T(E)} = \frac{1}{\lambda_\Lambda^{(\rm sh)}} -\int_{\vec{q}}^{\Lambda} \frac{1}{E+\rmi 0-2q^2},
\end{align}
where $2q^2=q^2/2M_{\rm r}$ with reduced mass $M_{\rm r}=M/2$, see, for instance, Ref. \cite{PhysRevA.65.022706,Levinsen2015} for a detailed discussion. In 3D the integration yields
\begin{align}
 \label{T3D} \frac{1}{T^{\rm 3D}(E)} = \frac{1}{8\pi}\Bigl(\frac{1}{a_{\rm 3D}}-\sqrt{-(E+\rmi 0)/2}\Bigr)
\end{align}
provided we choose
\begin{align}
 \label{lamLam3D} \frac{1}{\lambda_{{\rm 3D},\Lambda}^{(\rm sh)}} = -\frac{\Lambda}{4\pi^2} +\frac{1}{8\pi a_{\rm 3D}}.
\end{align}
In the 2D case we have
\begin{align}
 \nonumber \frac{1}{T^{\rm 2D}(E)} &= -\frac{1}{8\pi} \log\Bigl(-\frac{(E+\rmi 0)a_{\rm 2D}^2}{2}\Bigr) \\
 &= -\frac{1}{8\pi} \Bigl[\log\Bigl(\frac{Ea_{\rm 2D}^2}{2}\Bigr)-\rmi \pi\Bigr]
\end{align}
and
\begin{align}
 \label{2dini} \frac{1}{\lambda_{{\rm 2D},\Lambda}^{\rm (sh)}}  = -\frac{1}{8\pi}\log(\Lambda^2 a_{\rm 2D}^2).
\end{align}
These universal forms of the T-matrix for low-energy scattering in 3D and 2D serve as definitions of the scattering lengths $a_{\rm 3D}$ and $a_{\rm 2D}$.

We now consider the generalization of $T(E)$ when the $z$-direction is confined in a well of size $L$ with pbc. We then have the quantization $q_z\to k_n$, $n\in\mathbb{Z}$, for momenta in the $z$-direction, and
\begin{align}
 \int_{-\infty}^\infty \frac{\mbox{d}q_z}{2\pi} \to \sum_{n=-\infty}^\infty \frac{\Delta k_n}{2\pi} = \frac{1}{L}\sum_{n=-\infty}^\infty.
\end{align}
The zero-point energy vanishes for this particular choice of confinement. We write $A=-(E+\rmi 0)/2$ and find
\begin{align}
 \nonumber \frac{1}{T(E,L)} &= \frac{1}{\lambda_{{\rm 3D},\Lambda}^{(\rm sh)}} + \frac{1}{2L}\sum_n \int^{\Lambda'}\frac{\mbox{d}^2q}{(2\pi)^2} \frac{1}{k_n^2+q^2+A}\\
 \nonumber &= \frac{1}{\lambda_{{\rm 3D},\Lambda}^{(\rm sh)}} + \frac{1}{2}\int^{\Lambda'} \frac{\mbox{d}^2q}{(2\pi)^2} \frac{\frac{1}{2}+n_B(L\sqrt{q^2+A})}{\sqrt{q^2+A}}\\
 &= \frac{1}{\lambda_{{\rm 3D},\Lambda}^{(\rm sh)}} + \frac{\Lambda'-\sqrt{A}}{8\pi} - \frac{1}{4\pi L} \log(1-e^{-L\sqrt{A}})
\end{align}
with $n_B(x)=(e^x-1)^{-1}$. For $L=\infty$, the result is finite in the limit $E\to 0$. It reduces to the 3D expression (\ref{T3D}) provided we choose
\begin{align}
 \frac{1}{\lambda_{{\rm 3D},\Lambda}^{(\rm sh)}} = -\frac{\Lambda'}{8\pi}+\frac{1}{8\pi a_{\rm 3D}},\ \Lambda'= \frac{2}{\pi}\Lambda.
\end{align}
(Note that $\Lambda$ and $\Lambda'$ are UV cutoffs for 3D and 2D continuous momenta, respectively, and thus do not need to coincide.) More formally we define
\begin{align}
 \frac{1}{8\pi a_{\rm 3D}} := \lim_{E\to 0} \lim_{L\to \infty}\frac{1}{T(E,L)}.
\end{align}
We thus have
\begin{align}
\label{fff} \frac{1}{T(E,L)} &=\frac{1}{8\pi}\Bigl(\frac{1}{a_3}-\sqrt{A}\Bigr)-\frac{1}{4\pi L} \log(1-e^{-L\sqrt{A}}).
\end{align}
For $L<\infty$, the T-matrix develops a logarithmic IR-singularity as $E\to 0$. In particular, we find for the effective 2D-T-matrix
\begin{align}
 \nonumber \frac{1}{T_{\rm 2D}(E)} &:= \frac{L}{T(E,L)} \stackrel{A\to0}{\simeq} \frac{L}{8\pi a_{\rm 3D}} -\frac{1}{4\pi}\log(L\sqrt{A})\\
 \nonumber &=  \frac{L}{8\pi a_{\rm 3D}} -\frac{1}{4\pi}\log\Bigl(L\sqrt{-\frac{(E+\rmi 0)}{2}}\Bigr) \\
 \nonumber &\stackrel{!}{=} -\frac{1}{4\pi}\log\Bigl(a_{\rm 2D}\sqrt{-\frac{(E+\rmi 0)}{2}}\Bigr)\\
 \label{logE} &= -\frac{1}{8\pi} \log\Bigl(- \frac{(E+\rmi 0)a_{\rm 2D}^2}{2}\Bigr).
\end{align}
We observe the effective 2D scattering length of the confined system to be
\begin{align}
 a_{\rm 2D}^{\rm (pbc)}(L) = L \exp\Bigl\{-\frac{1}{2}\frac{L}{a_{\rm 3D}}\Bigr\}.
\end{align}
This result agrees with the FRG finding in Eq. (\ref{2d23}).

\section{Discussion}\label{SecDiss}

In this work we have studied the dimensional crossover of non-relativistic bosons when confining a transverse spatial dimension. Although the derived set of flow equations is applicable to crossovers from $(d+1)$ to $d$ dimensions, our main focus has been on the crossover from 3D to 2D. The motivation for this is the intriguing nature of the superfluid transition in this case and the considerable number of recent cold atom experiments on this topic. Especially, we wish to contribute to a few open theoretical challenges in the interpretation of the data of these as outlined below.

The present approach may, however, be extended to describe further dimensional crossovers in condensed matter systems. Compactifications from 2D to 1D, or 3D to 1D, are highly exciting from the point of view of quantum phase transitions. Indeed, phases with quasi long range order may survive in 1D at zero temperature and are conceptually close to the finite temperature superfluid phase in the 2D Bose gas, see for instance Ref. \cite{PhysRevB.77.064504} for an FRG perspective on this aspect, and Ref. \cite{PhysRevA.68.043603} for a mean-field study of the correlations in the dimensional crossover from 3D to 2D (1D) at $T>0$ ($T=0$). Since the confinement only affects spatial momenta, the characteristic frequency dependence which governs quantum critical phenomena will be resolved correctly with the FRG in the dimensional crossover if this is the case for the system without compact dimensions. The nature of inhomogeneous phases of fermion pairs in dimensional crossovers have been discussed in Refs. \cite{PhysRevA.85.051607,2015arXiv150803352D}.

As an effective description of bosonic degrees of freedom, the finite temperature 2D Bose gas and its BKT transition may also be applied to the pseudogap phase of underdoped cuprates \cite{RevModPhys.78.17}. The inherent non-relativistic nature of complex bosonic degrees of freedom is also found in open or non-equilibrium quantum systems such as exciton-polariton condensates in semiconductor microcavities \cite{RevModPhys.85.299,2015arXiv151200637S}. The Keldysh field theory techniques required to describe such systems can be incorporated into the FRG framework, see Ref. \cite{Berges:2012ty} for an introduction.

In our analysis of the superfluid phase transition we have performed the dimensional crossover in two variants: for fixed $a_{\rm 3D}$ and for fixed $a_{\rm 2D}$. Each procedure addresses a somewhat different question. When fixing $a_{\rm 3D}$ we follow the typical cold atom experimental protocol of setting the value of $a_{\rm 3D}$ by means of an external magnetic field $B$ and then decreasing the length $L$ to eventually arrive at a 2D system. Importantly, the size of $L$ is typically not easily tunable in experiment but rather fixed to a particular value. We can then reach a large class of values for $a_{\rm 2D}$ by means of formula (\ref{2d23}) by changing $a_{\rm 3D}(B)$. In view of Fig. \ref{fig:Tc_vs_L}, where $g_{\rm 3D}\sqrt{\mu}=0.025$, one experiment corresponds to one certain point $L\sqrt{\mu}$ on the abscissa. Through Eq. (\ref{abcd}) this corresponds to a certain value of $g_{\rm 2D}$. If one were to perform the experiment with a different value of $L$, one would simulate a different Bose gas (with different $g_{\rm 2D}$) for the same value of $a_{\rm 3D}$. In order to plan and devise an experimental setup it is crucial to know the effective $a_{\rm 2D}$ and $g_{\rm 2D}$ which can be reached for a given $L$, and the formulas (\ref{2d24}) and (\ref{2d26}) are abundantly applied in the experimental cold atoms literature. 

On the other hand, in view of attempting to quantum simulate 2D systems, we may ask: Given $L\neq 0$, can we be sure that we are measuring the critical temperature of the 2D system and not that of its quasi-2D relatives? By fixing $a_{\rm 2D}$ (instead of $a_{\rm 3D}$) for small $L$ in the dimensional crossover we confirm with the study in App \ref{AppTca2D} that the critical temperature is enhanced by the residual influence of the third dimension. This is not too surprising as we expect the possibility of Bose condensation in $d>2$ to largely increase the critical temperature. Furthermore, we found that the true 2D critical temperature is realized for a considerable range of $(L/a_{\rm 2D})$-values.

Our discussion of the low-energy T-matrix to describe the scattering properties of the Bose gas was mostly inspired by the aim to benchmark the relation (\ref{2d23}) between the 3D and 2D scattering lengths obtained with the FRG. However, given an exact or approximate expression for the T-matrix of the confined system, an obstacle of analysing current cold atom experiments can be resolved: At large densities -- especially for fermionic system, which do not differ from the bosonic counterparts in their scattering properties -- the typical momenta of particles are not necessarily small. Hence one needs to restrict the experiment to low densities, or the scattering cross sections and effective coupling constants need to be evaluated at a finite momentum \cite{PhysRevLett.112.045301,Dyke2014,Fenech2015,Boettcher2015}. This may effectively be accomplished with the energy- and momentum-resolved T-matrix of the confined system with its particular discrete mode spectrum such as in Eq. (\ref{fff}). To resolve this mode spectrum, one may actually restrict to phenomenological collective mode spectra such as in Refs. \cite{PhysRevA.68.043610,PhysRevLett.93.040402,Boettcher:2011iq}. Complementary approaches to obtain the scattering properties inside a medium from the vacuum T-matrix, however for a true 2D system, have been successfully applied, for instance, to the Fermi polaron problem \cite{PhysRevA.85.021602}.

Through confining the gas by means of a potential well with pbc, the challenging task of determining the interacting ground state of the system became particularly simple: The ground state is still homogeneous in space and given by the minimum of the effective potential $U_{k\to 0}(\phi^2)$. However, as discussed at the end of Sec. \ref{SecModel}, the situation becomes more complicated already for a box potential with infinite walls. Whereas in a Hamiltonian formulation the ground states of the interacting system are required for computing expectation values of observables, the interacting ground state enters the functional integral formulation by the need to evaluate the effective action $\Gamma[\phi]$ at its minimum configuration $\phi_0(\tau,\vec{x})$. The latter, however, will not be constant in space for a trapped system, i.e., we are dealing with an inhomogeneous ground state. Thus, strictly speaking, the ground state can no longer be found by minimizing the effective potential $U_{k\to 0}(\phi^2)$ evaluated for constant fields. Still one might define a suitable modification of the effective potential which captures the energy of the lowest lying excitations, which might be modelled in a first approximation by the non-interacting states or deformations thereof.

At last we discuss the relevance of our findings for trapping potentials different from a well with pbc. The main difference when trapping the system in a box potential or a harmonic trap consists (i) in the fact that the ground states of the latter are inhomogeneous and (ii) that the related ground state energies $E_0$ are nonzero. We already addressed the first point in the previous paragraph. In fact, it should constitute the main technical difficulty in the computation of different confinement scenarios with the FRG.

To understand the role of (ii) note that the emergent 2D scattering physics relied on a logarithmic singularity of the effective 2D-T-matrix as $E\to 0$, see Eq. (\ref{logE}). This divergence allows us to read off the effective 2D scattering length. For a system with nonvanishing $E_0$, the logarithmic divergence will appear for $\Delta E\to 0$ where $E=E_0+\Delta E$. This is true for harmonic confinement \cite{PhysRevA.64.012706,Levinsen2015}, and a preliminary calculation for the T-matrix in a box suggests that the same pattern appears, with a formula relating $a_{\rm 3D}$ and $a_{\rm 2D}$ analogous to Eq. (\ref{2d23}). Consequently, the emergent 2D physics appear in the effective 2D continuum of states just above the zero point energy, see Fig. \ref{fig:energy_spectrum}. Subtracting the shift of energies due to $E_0$, e.g. by a chemical potential or by evaluating the ``effective potential'' at a nonzero frequency, the remaining field theory describing scattering and the superfluid phase transition will be very similar to the one considered here with pbc. In fact, we have shown that Eqs. (\ref{2d23}) and (\ref{2d25b}) are in one-to-one correspondence to the results for harmonic confinement with $\ell_z$ once we identify $\ell_z^{\rm eff}=L/\sqrt{2\pi}$ as the effective oscillator length of the potential well.

\begin{center}
 \textbf{Acknowledgements}
\end{center}

\noindent We thank Jan M. Pawlowski, Marcelo T. Yamashita, and Nikolaj T. Zinner for inspiring discussions and helpful comments. This work has been supported by the ERC through the Advanced Grant No. 290623. S. L. acknowledges financial support by the Heidelberg Graduate School of Fundamental Physics. I. B. acknowledges funding by the DFG under Grant No. BO 4640/1-1.

\begin{appendix}

\section{Superfluid transition in the 2D limit}

\subsection{Critical temperature in two dimensions}\label{AppTc2D}
The limiting formula (\ref{stcrit4}) for the critical temperature of a 2D superfluid Bose gas can be studied independently of the details of the dimensional crossover. For this purpose we set $\Lambda/\sqrt{\mu}=10^3$ as in the main text, and vary the values of $t_{\rm f}=\log(k_{\rm f}/\Lambda)$ and $\lambda_{{\rm 2D},\Lambda}$. We define
\begin{align}
 \label{appTc1} g_{\rm 2D} =\lambda_{\rm 2D, k=\sqrt{T}}
\end{align}
as in Eq. (\ref{ab}).

\begin{figure}[t]\centering
\includegraphics[width=8cm]{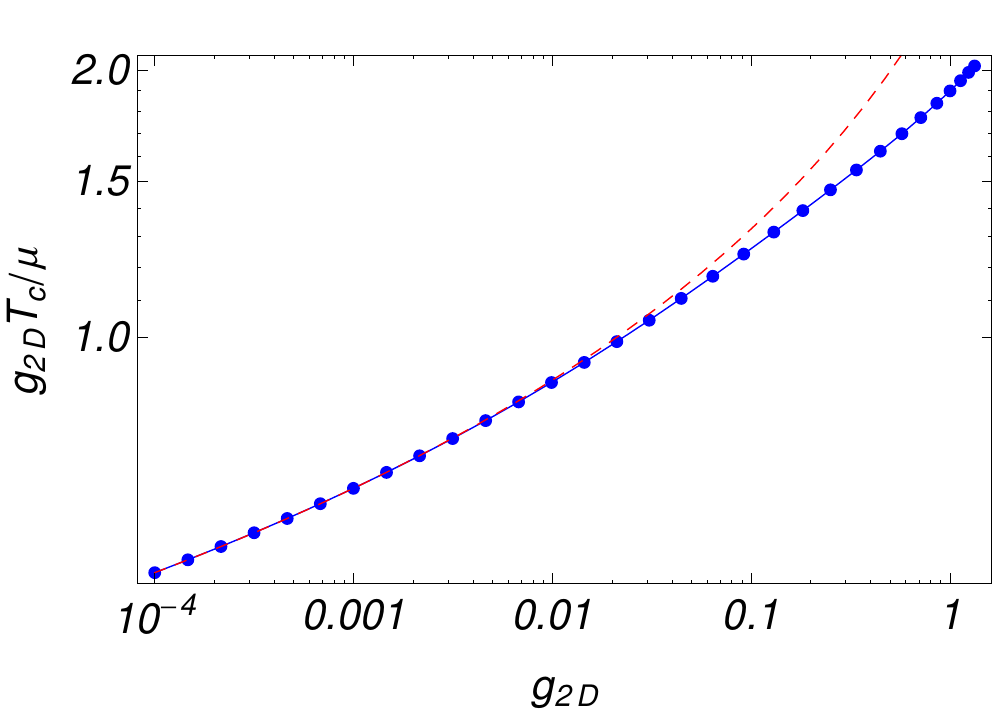}
\caption{Critical temperature $T_{\rm c}/\mu$ in 2D as a function of $g_{\rm 2D}$ over four orders of magnitude. For better comparability with Eq. (\ref{appTc2}) we have multiplied the function by $g_{\rm 2D}$. The blue points and solid line constitute the FRG result for $t_{\rm f}=-10$ (as used in the main text), whereas the red dashed line corresponds to the asymptotic form of Eq. (\ref{appTc2}) for small $g_{\rm 2D}$. For the dashed line we use the fit result from Tab. \ref{AppTcTab}.}
\label{FigTcg}
\end{figure}

For $t_{\rm f}=-10$, as employed in the main text, we find the $g_{\rm 2D}$-dependence of $T_{\rm c}/\mu$ to be in good agreement with the asymptotic form of Eq. (\ref{stcrit4}), i.e.,
\begin{align}
 \label{appTc2} \frac{T_{\rm c}}{\mu} = \frac{2\pi}{g_{\rm 2D}} \frac{A}{\log(2\xi_\mu/g_{\rm 2D})},
\end{align}
when choosing $A=0.99$ and $\xi_{\mu}=6.0$. We visualize the coupling dependence of the critical temperature in Fig. \ref{FigTcg}. Upon varying $t_{\rm f}$ in a reasonable range, we consistently find scaling according to Eq. (\ref{appTc2}) with only mild variations of the parameters $A$ and $\xi_\mu$, which we collect in Tab. \ref{AppTcTab}. The corresponding values have been obtained from a fit of $T_{\rm c}/\mu$ to Eq. (\ref{appTc2}) in the interval $g_{\rm 2D}\in[10^{-4},0.01]$.
\begin{table}[h]
\begin{tabular}{|c||c|c|c|c|c|}
\hline $t_{\rm f}$ & -8 & -9 & -10 & -11 & -12\\ 
\hline\hline $A$ & 0.94 & 0.98 & 0.99 & 0.99 & 0.99\\
\hline $\xi_\mu$ & 5.1 & 5.3 & 6.0 & 6.6 & 7.2\\
\hline
\end{tabular} 
\label{AppTcTab}
\caption{Fitting parameters obtained from matching the critical temperature $T_{\rm c}/\mu$ for different values of $t_{\rm f}=\log(k_{\rm f}/\Lambda)$ to Eq. (\ref{appTc2}) in the interval $g_{\rm 2D}\in[10^{-4},0.01]$.}
\end{table}

\subsection{Dimensional crossover for fixed $a_{\rm 2D}$}\label{AppTca2D}

In order to perform the dimensional crossover for fixed $a_{\rm 2D}$ we need to choose the initial conditions of the flow rather careful. Here we collect the technical details and parameters which allow us to conclude that the critical temperature of the 2D gas is enhanced by the influence of a sufficiently large third dimension.

First note that Eq. (\ref{2d12}) implies that $a_{\rm 3D}$ cannot be arbitrary large within our model with pointlike interactions. Indeed, for fixed $\Lambda$, the scattering length is bounded from above according to $a_{\rm 3D}< a_{\rm max}=(\frac{4\Lambda}{3\pi})^{-1}=2.36\Lambda^{-1}$ \cite{PhysRevA.77.053603}. In particular, for $\Lambda\to \infty$ we have $a_{\rm 3D}\to 0$, meaning that the pointlike approximation cannot describe the interacting system up to arbitrarily high energies. For fixed $L$ this implies $a_{\rm 2D}< L e^{-\frac{1}{2}(L/a_{\rm max})}$ due to Eq. (\ref{2d23}). Thus, in typical experiments with fixed $L$, the range of accessible $a_{\rm 2D}$ is also limited from above.

\begin{figure}[t]\centering
\includegraphics[width=8cm]{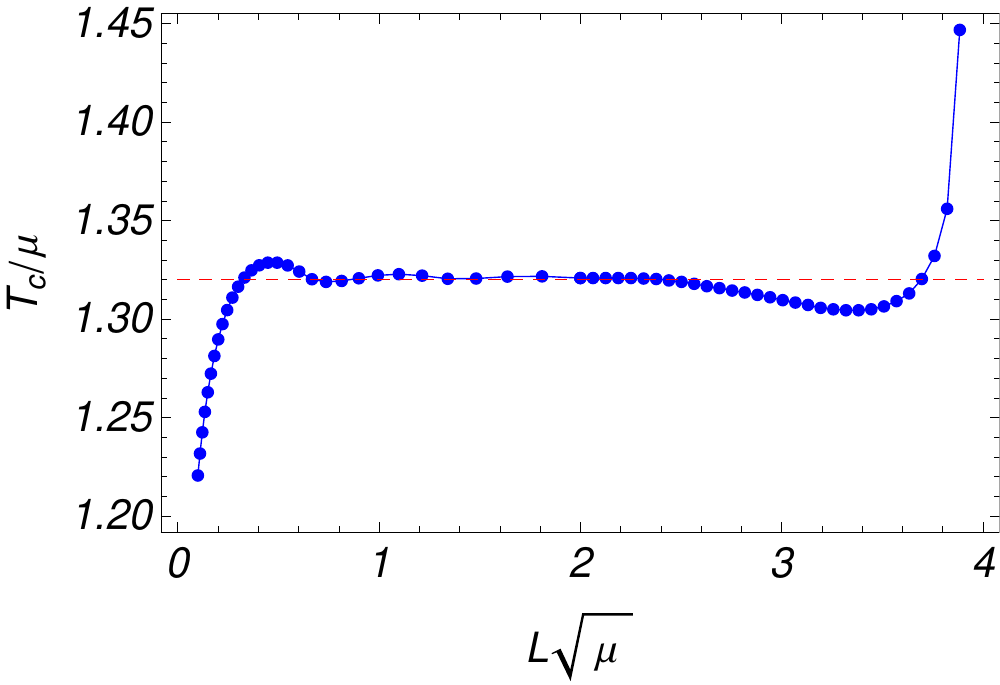}
\caption{Dimensional crossover of the superfluid critical temperature $T_{\rm c}$ for a fixed value of $a_{\rm 2D}$. For this figure we use the parameters as described in this appendix. One observes three characteristic regimes for the behavior of the critical temperature: (i) For small $0.1\leq L\sqrt{\mu}\leq 1$, the transition temperature increases. However, since we have chosen $\Lambda/\sqrt{\mu}=10$ here, this region corresponds to $1\leq L\Lambda\leq 10$ and thus violates the criterion $L\gg\Lambda^{-1}$. Thus it is not of relevance for this discussion. (ii) For intermediate $L\sqrt{\mu}$, the critical temperature settles at $T_{\rm c}/\mu=1.32$ (horizontal dashed line), which is the transition temperature for the corresponding 2D gas with the same value of $a_{\rm 2D}$. In this region, which corresponds to $L/a_{\rm 2D} =(1-3)\times 10^3$ in this particular case, the gas is truly two-dimensional. (iii) For larger $L$, we find an \emph{increase} of the critical temperature. Note that the crossover scale $L\sqrt{\mu}\approx 3$ between the 2D and the quasi-2D regime is still much larger than $T^{-1/2}$ and $\mu^{-1/2}$. The slight dimple in $T_{\rm c}/\mu$ before the rising seems to root in the same oscillatory crossover behavior as also found for small $L$, and we do not attribute any deep physical meaning to it at this point.}
\label{FigTca2D}
\end{figure}

On the other hand, one may fix $a_{\rm 2D}$ and $\Lambda$, and determine the possible values of $L$ which realize this particular $a_{\rm 2D}$. This leads us to considering the function $f(x)=4\pi a_{\rm 2D}/\lambda_\Lambda^{\rm 3D}=\frac{\log(x)}{x}-\frac{2}{3\pi}(\Lambda a_{\rm 2D})$ with $x=L/a_{\rm 2D}$. A stable solution requires $f(x)$ to be positive, leading to a stability interval $[L_{\rm min},L_{\rm max}]$ of possible $L$-values. From Eqs. (\ref{2dini}) and (\ref{t3b}) we see that, disregarding some modifications due to the regularization procedure, we have $\Lambda a_{\rm 2D}<1$. For instance, choosing $\Lambda a_{\rm 2D}=0.1\ (0.01)$ we arrive at $x_{\rm min}\simeq1.02\ (1.002)$ and $x_{\rm max}\simeq260\ (3900)$. The corresponding value of $L_{\rm min}$, however, will typically be too small as we also need to satisfy $L\Lambda\gg 1$.

For our discussion of the dimensional crossover at fixed $a_{\rm 2D}$ we employ $\mu=1$, $a_{\rm 2D}=10^{-3}$, $\Lambda=10$, $k_{\rm f}=10^{-2}$, and vary $L$ within the boundaries allowed by these constraints. We choose $L_{\rm min}=\Lambda^{-1}$ from $L\Lambda>1$ and $L_{\rm max}=3885a_{\rm 3D}$ from the positivity of $f(x)$. The critical temperature of the pure 2D gas for this choice of parameters is $T_{\rm c}/\mu=1.32$. The resulting crossover of the critical temperature is shown in Fig. \ref{FigTca2D}.

\section{Flow Equations}\label{AppFlow}

In this appendix we derive the flow equation for the Bose gas in the dimensional crossover within the derivative expansion for the effective action. In this approach, the kinetic energy of spatial excitations is approximated by a dispersion relation
\begin{align}
E_{\vec{q}} = \frac{1}{\bar{Z}(k)}\Bigl(A(k) (q_1^2+\dots+q_d^2)+A_{\rm z}(k) q_{d+1}^2\Bigr)
\end{align}
with $k$-dependent running couplings $\bar{Z}(k)$, $A(k)$, and $A_{\rm z}(k)$. The distinction between $A$ and $A_{\rm z}$ is subleading for the physics under consideration, as can be understood as follows: For large $k$ we have $A(k)=A_{\rm z}(k)$ and the dispersion is effectively isotropic. Deviations between $A_{\rm z}$ and $A$ appear for $k\lesssim L^{-1}$. In this regime, however, excitations with nonzero $q_{d+1}^2=k_n^2=(\frac{2\pi n}{L})^2>0$ are disfavoured by an energetic penalty $\sim L^{-2}$. Accordingly, as long as the prefactor $A_{\rm z}$ remains of order unity (i.e. it does not become exceptionally small), transverse excitations $\sim A_{\rm z}k_n^2$ decouple from the RG flow, and the precise value of the prefactor $A_{\rm z}$ is unimportant.

In the following we first derive the flow equations for the isotropic case with $A=A_{\rm z}$. This approximation is applied throughout the main text. Afterwards we consider the more involved anisotropic case with $A\neq A_{\rm z}$. We explicitly show that this distinction is unimportant for the observables considered here. In particular, as can be seen in Fig. \ref{FigAniso}, the ratio $\xi=A_{\rm z}/A$ remains of order unity, thus validating the above argument.

\subsection{Isotropic derivative exapnsion}

Our Ansatz for the effective average action is given by a derivative expansion
\begin{align}
 \label{FuncFlow} \Gamma_k\left[\bar{\phi}^*,\bar{\phi}\right]=\int_X\Bigl(\bar{\phi}^*\left(\bar{Z}_k\partial_\tau-A_k\nabla^2\right)\bar{\phi}+\bar{U}_k\left(\bar{\phi}^*\bar{\phi}\right)\Bigr).
\end{align}
By introducing renormalized fields $\phi=A_k^{1/2}\bar{\phi}$ we can express the effective average action as
\begin{align}
\Gamma_k\left[\phi^*,\phi\right]=\int_X\Bigl(\phi^*\left(Z_k\partial_\tau-\nabla^2\right)\phi+U_k\left(\phi^*\phi\right)\Bigr)
\end{align}
with wavefunction renormalization $Z_k=\bar{Z}_k/A_k$ (see Eq. (\ref{m3})) and the effective average potential $U_k(\rho)=\bar{U}_k(\bar{\rho})$. For a detailed discussion of obtaining the projected flow equations for $\bar{Z}_k$, $A_k$, and $\bar{U}_k$ from the functional flow equation in Eq. (\ref{FuncFlow}), see Apps. A-C of Ref. \cite{PhysRevA.89.053630}.

The flow equation for the effective potential is given by
\begin{align}
\label{fe1}\dot{\bar{U}}_k(\bar{\rho})=\frac{1}{2}\text{tr}\int_Q\bar{G}_k(Q)\dot{R}_k(Q)
\end{align}
with the regularized propagator
\begin{align}
\label{fe2}\bar{G}_k(Q)=\frac{1}{\text{det}_Q}
\left(
\begin{array}{cc}
	p^q+\bar{U}' & \bar{Z}_kq_0\\
	-\bar{Z}_kq_0 & p^q+\bar{U}'+2\bar{\rho}\bar{U}''
\end{array}
\right),
\end{align}
where $Q=(q_0,\vec{q},k_n)$, $\vec{q}=(q_1,\dots,q_d)$ denotes the momentum vector in the noncompact dimensions of space, and $k_n$ the component of the discrete momentum modes due to the transverse confinement. Below we specialize to the case that $d=2$. Further,
\begin{align}
\label{fe3}\text{det}_Q&=(p^q+\bar{U}'+2\bar{\rho}\bar{U}'')(p^q+\bar{U}')+(\bar{Z}_kq_0)^2,\\
p^q&=A_k(\vec{q}^2+k_n^2)+R_k(Q),
\end{align}
where a prime denotes a derivative with respect to $\bar{\rho}$ and a dot denotes $\partial_t=k\partial_k$.
In this calculation we use the Litim-type regulator  \cite{Litim:2001fd,PhysRevB.77.064504,Litim:2000ci}
\begin{align}
\label{fe4}R_k(Q)&=A_k(k^2-\vec{q}^2-k_n^2)\theta(k^2-\vec{q}^2-k_n^2)\\
\dot{R}_k(Q)&=A_k(2k^2-\eta(k^2-\vec{q}^2-k_n^2))\theta(k^2-\vec{q}^2-k_n^2),
\end{align}
where we define the anomalous dimension by $\eta=-\dot{A}_k/A_k$. The integration in Eq. (\ref{fe1}) consists of a $q_0$-integral, which is replaced by a sum over Matsubara frequencies $\omega_n$ for non-zero temperatures, and an integration over spatial dimensions. Therefore, we have
\begin{align}
 \label{fe5}\int_Q=T\sum_{\omega_n}\frac{1}{L}\sum_{k_n}\int \frac{d^{d}q}{(2\pi)^d}.
\end{align}

The flow equation for the effective potential can be written in the form
\begin{align} 
\label{fe6}
\nonumber\dot{\bar{U}}_k&=\int_Q\frac{p^q+\bar{U}'+\bar{\rho}\bar{U}''}{\text{det}_Q}\dot{R}_k\\
\nonumber&=T\sum_{\omega_n}\frac{1}{L}\sum_{k_n}4v_d\int_0^\infty \text{d}qq^{d-1}\frac{p^q+\bar{U}'+\bar{\rho}\bar{U}''}{\text{det}_Q}\\
&\times A_k\left(2k^2-\eta(k^2-q^2-k_n^2)\right)\theta\left(k^2-q^2-k_n^2\right).
\end{align}
Here we have already performed the angular integration of the $d$-dimensional integral, which gives a factor of $4v_d=[2^{d-1}\pi^{d/2}\Gamma(d/2)]^{-1}$. The overall $\theta$-function limits the integration to the region $q<\sqrt{k^2-k_n^2}$ and we may replace $p^q\rightarrow A_k k^2$ in the integrand.
We switch to renormalized quantities $Z_k=\bar{Z}_k/A_k$ and $\rho=A_k\bar{\rho}$ and find
\begin{align}
\label{fe7}
\nonumber\dot{\bar{U}}_k&=4v_d\left(T\sum_{\omega_n}\frac{k^2+U'+\rho U''}{(k^2+U')(k^2+U'+2\rho U'')+(Z_k\omega_n)^2}\right)\\
\nonumber&\times \frac{1}{L}\sum_{k_n}\int_0^{\sqrt{k^2-k_n^2}}\text{d}qq^{d-1}\left(2k^2-\eta(k^2-q^2-k_n^2)\right)\\
&\times \theta(k^2-q^2-k_n^2)
\end{align}
For the $q$-integration we obtain analytically
\begin{align}
\label{fe8}
\nonumber&\int_0^{\sqrt{k^2-k_n^2}}\text{d}qq^{d-1}\left(2k^2-\eta(k^2-q^2-k_n^2)\right)\theta(k^2-q^2-k_n^2)\\
&=\frac{2}{d}k^{2+d}(1-(k_n/k)^2)^{d/2}\left(1-\eta\frac{1-(k_n/k)^2}{d+2}\right)\theta(k^2-k_n^2).
\end{align}
The temperature-dependent part evaluates to
\begin{align}
\label{fe9}
\nonumber&T\sum_{\omega_n}\frac{k^2+U'+\rho U''}{(k^2+U')(k^2+U'+2\rho U'')+(Z_k\omega_n)^2}\\
\nonumber=&\frac{1}{2Z_k}\left(\sqrt{\frac{1+w_1}{1+w_2}}+\sqrt{\frac{1+w_2}{1+w_1}}\right)\\
\times&\left(\frac{1}{2}+N_B\left(\frac{k^2\sqrt{(1+w_1)(1+w_2)}}{Z_k},\right)\right),
\end{align}
where we introduced the dimensionless quantities $w_1=U'/k^2$, $w_2=(U'+2\rho U'')/k^2$ and the Bose function $N_B(x)=(e^{x/T}-1)^{-1}$. The flow for the effective potential therefore reads
\begin{align}
\label{fe10}
\nonumber\dot{\bar{U}}_k&=\frac{4v_d k^{3+d}}{dZ_k}G_1(T)F(\tilde{L}),\\
\nonumber G_1(T)&=\left(\sqrt{\frac{1+w_1}{1+w_2}}+\sqrt{\frac{1+w_2}{1+w_1}}\right)\left(\frac{1}{2}+N_B\right),\\
\nonumber F(\tilde{L})&=\frac{1}{Lk}\sum_{k_n}(1-(k_n/k)^2)^{d/2}\\
&\times\left(1-\eta\frac{1-(k_n/k)^2}{d+2}\right)\theta(k^2-k_n^2)
\end{align}
with $\tilde{L}=Lk$.

We now specialize to two continuous dimensions. Further, we compactify on a torus, thus $k_n=2\pi n/L$, $n\in\mathbb{Z}$. The $\theta$-function limits the range of summation in the crossover-function to $\lvert k_n\rvert=\lvert\frac{2\pi}{L}n\rvert<k$, or, equivalently, $\lvert n\rvert <\frac{\tilde{L}}{2\pi}$. With $N=\left\lfloor \frac{\tilde{L}}{2\pi}\right\rfloor$ we can write
\begin{align}
\label{fe11}
\nonumber F_{\rm pbc}(\tilde{L})&=\frac{1}{\tilde{L}}\left[\left(1-\frac{\eta}{4}\right)\sum_{n=-N}^{n=N}1\right.\\
&\left.-\left(1-\frac{\eta}{2}\right)\sum_{n=-N}^{n=N}\frac{k_n^2}{k^2}-\frac{\eta}{4}\sum_{n=-N}^{n=N}\frac{k_n^4}{k^4}\right].
\end{align}
Together with
\begin{align}\label{fe12}
\nonumber\sum_{n=-N}^N\frac{k_n^2}{k^2}&=2\left(\frac{2\pi}{\tilde{L}}\right)^2\sum_{n=1}^Nn^2,\\
\sum_{n=-N}^N\frac{k_n^4}{k^4}&=2\left(\frac{2\pi}{\tilde{L}}\right)^4\sum_{n=1}^Nn^4
\end{align}
and
\begin{align}
\label{fe13}
\nonumber\sum_{n=-N}^{n=N}1&=1+2N,\\
\nonumber\sum_{n=1}^Nn^2&=\frac{1}{6}N(1+N)(1+2N),\\
\sum_{n=1}^Nn^4&=\frac{1}{30}N(1+N)(1+2N)(-1+3N+3N^2)
\end{align}
the crossover-function evaluates to
\begin{align}
\label{fe14}
\nonumber F_{\rm pbc}(\tilde{L})&=\frac{2N+1}{\tilde{L}}\left[1-\frac{\eta}{4}-\frac{1}{\tilde{L}^2}\left(1-\frac{\eta}{2}\right)\frac{4\pi^2}{3}N(N+1)\right.\\
&\left.-\frac{\eta}{\tilde{L}^4}\frac{4\pi^4}{15}N(N+1)(-1+3N+3N^2)\right].
\end{align}
The flow for the effective potential with $v_2=1/8\pi$ and $\dot{U}_k=\dot{\bar{U}}_k+\eta\rho U'_k$ reads 
\begin{align}
\label{fe15}
\dot{U}_k=\eta\rho U'_k +\frac{k^5}{4\pi Z_k}G_1(T)F(\tilde{L}).
\end{align}

We use the derivative projection to obtain the flow equations for the wavefunction renormalizations $\bar{Z}_k$ and $A_k$ defined via
\begin{align}
\label{fe16}
\nonumber\partial_t\bar{Z}_k&=-\frac{\partial}{\partial p_0}\dot{\bar{G}}_{k,12}^{-1}(p_0,0)\bigg|_{p_0=0}\\
\partial_tA_k&=\frac{\partial}{\partial p^2}\dot{\bar{G}}_{k,22}^{-1}(0,p^2)\bigg|_{p=0},
\end{align}
where $G_{k,ij}^{-1}(p_0,p^2)=\delta^2\Gamma/(\delta\phi_i\delta\phi_j)|_{\rho_0}$ denotes the second functional derivative.
For the flow of $\bar{Z}_k$ we arrive at
\begin{align}
\label{fe17}
\nonumber \bar{Z}_k^{-1}\partial_t\bar{Z}_k&=4\rho U''^2\int_Q\frac{k^4-2\rho U''(k^2+\rho U'')+(Z_kq_0)^2}{\left(k^2(k^2+2\rho U'')+(Z_kq_0)^2\right)^3}\\
&\times\left(2k^2-\eta(k^2-q^2-k_n^2)\right)\theta(k^2-q^2-k_n^2).
\end{align}
Again we can evaluate
\begin{align}
\label{fe18}
\nonumber&\frac{1}{L}\sum_{k_n}\int\frac{\text{d}^dq}{(2\pi)^d}\left(2k^2-\eta(k^2-q^2-k_n^2)\right)\theta(k^2-q^2-k_n^2)\\
&=4v_d\frac{2}{d}k^{3+d}F(\tilde{L})
\end{align}
with $F(\tilde{L})$ as defined in Eq. (\ref{fe10}).
Together with $U''=\lambda$ and $\bar{Z}_k=Z_kA_k$ we obtain
\begin{align}
\label{fe19}
\partial_tZ_k=\eta Z_k-Z_k4\rho\lambda^2\frac{8v_dk^{3+d}}{d}\frac{1}{2Z_kk^6}G_2(T)F(\tilde{L}),
\end{align}
where we defined the temperature-dependent function
\begin{align}
\label{fe21}
G_2(T)=&\frac{1}{(1+w_2)^{3/2}}\left(\frac{1}{2}+N_B(c)-cN'_B(c)\right)\\
 \nonumber &-\frac{3}{8}\frac{w_2(4+w_2)}{(1+w_2)^{5/2}}\\
\nonumber & \times\left(\frac{1}{2}+N_B(c)-cN'_B(c)+\frac{c^2}{3}N''_B(c)\right),
\end{align}
with $c=k^2\sqrt{1+w_2}/Z_k$ and $N_B(c)=(e^{c/T}-1)^{-1}$. Specifying to $d=2$ noncompact dimensions we arrive at
\begin{align}
\label{fe22}
\partial_tZ_k=\eta Z_k-\frac{\rho\lambda^2}{\pi k}G_2(T)F(\tilde{L}).
\end{align}

For the flow equation for $A_k$ we need to take a derivative with respect to $\left|\vec{p}\right|^2$. We consider the external momentum as a two-dimensional vector $\vec{p}=(p_x,p_y)$. We then have with $x=(\vec{q}\cdot\vec{p})/(qp)$ and $\eta=-(\partial_tA_k)/A_k$
\begin{align}
\label{fe31}
\nonumber\eta&=8\rho\lambda^2\left(T\sum_{\omega_n}\frac{1}{\text{det}_Q^2}\right)\frac{1}{L}\sum_{k_n}\int\frac{\text{d}^2q}{(2\pi)^2}q^2x^2\delta(k^2-q^2-k_n^2)\\
&\times (2k^2-\eta(k^2-q^2-k_n^2))\theta(k^2-q^2-k_n^2).
\end{align}
Angle integration over $x$ yields another factor of $1/2$. We have
\begin{align}
\label{fe32}
\nonumber\eta&=8\rho\lambda^2\left(T\sum_{\omega_n}\frac{1}{\text{det}_Q^2}\right)\frac{1}{L}\sum_{k_n}4v_2\frac{1}{2}\int\text{d}qqq^2\\
&\times(2k^2-\eta(k^2-q^2-k_n^2))\delta(k^2-q^2-k_n^2)\theta(k^2-q^2-k_n^2).
\end{align}
We further obtain
\begin{align}
\label{fe33}
\nonumber\eta&=\frac{32\rho\lambda^2v_2}{2}\left(T\sum_{\omega_n}\frac{1}{\text{det}_Q^2}\right)\frac{1}{L}\sum_{k_n}\\
&\times\frac{1}{2}\int\text{d}qq^3(2k^2-\eta(k^2-q^2-k_n^2))\delta(k^2-q^2-k_n^2).
\end{align}
By reformulating the $\delta$-function
\begin{align}
\label{fe34}
\nonumber\delta(k^2-k_n^2-q^2)&=\frac{1}{2\sqrt{k^2-k_n^2}}\left(\delta(\sqrt{k^2-k_n^2}-q)\right.\\
&\left.+\delta(\sqrt{k^2-k_n^2}+q)\right)
\end{align}
and using that the second $\delta$-function gives no contribution to the integral, we get
\begin{align}
\label{fe35}
\nonumber\eta&=\frac{32\rho\lambda^2v_2}{2}\left(T\sum_{\omega_n}\frac{1}{\text{det}_Q^2}\right)\frac{1}{L}\sum_{k_n}\\
\nonumber&\times\frac{1}{2}\frac{1}{2}(k^2-k_n^2)^{-1/2}(k^2-k_n^2)^{3/2}2k^2\\
\nonumber&=\frac{32\rho\lambda^2v_2}{2}\left(T\sum_{\omega_n}\frac{1}{\text{det}_Q^2}\right)\frac{1}{L}\sum_{k_n}\\
&\times\frac{1}{2}k^2(k^2-k_n^2).
\end{align}
We find
\begin{align}
\label{fe36}
\eta=\frac{32\rho\lambda^2}{32\pi }\left(T\sum_{\omega_n}\frac{1}{\text{det}_Q^2}\right)\frac{k^4}{L}\sum_{k_n}\left(1-(k_n/k)^2\right).
\end{align}
And finally
\begin{align}
\label{fe37}
\nonumber\eta&=\frac{\rho\lambda^2}{\pi}k^5\left(T\sum_{\omega_n}\frac{1}{\text{det}_Q^2}\right)\frac{1}{\tilde{L}}\sum_{k_n}\left(1-(k_n/k)^2\right)\\
&=\frac{\rho\lambda^2}{\pi}k^5\left(T\sum_{\omega_n}\frac{1}{\text{det}_Q^2}\right)F_0(\tilde{L}),
\end{align}
where $F_0(\tilde{L})$ is the crossover-function with $\eta$ set equal to zero, i.e.
\begin{align}
\label{fe38}
F_{0}(\tilde{L})&=\frac{2N+1}{\tilde{L}}\left[1-\frac{1}{\tilde{L}^2}\frac{4\pi^2}{3}N(N+1)\right].
\end{align}
The Matsubara summation yields
\begin{align}
\label{fe39} T\sum_{\omega_n}\frac{1}{\text{det}_Q^2}&=\frac{1}{2Z_kk^6}G_3(T).
\end{align}
with
\begin{align}
 G_3(T) = \frac{1}{(1+w_2)^{3/2}}\Bigl(\frac{1}{2}+N_B(c)-cN_B'(c)\Bigr),
\end{align}
with $c=k^2(1+w_2)^{1/2}/Z_k$ and $N_B(c)=(e^{c/T}-1)^{-1}$ as before. The anomalous dimension then reads
\begin{align}
\label{fe40}
\eta=\frac{\rho_0\lambda^2}{2\pi Z_kk}G_3(T)F_0(\tilde{L}).
\end{align}
This completes the derivation of the flow equations.

\subsection{Anisotropic derivative expansion}
In the following we consider the anisotropic extension of Eq. (\ref{FuncFlow}) given by
\begin{align}
 \Gamma_k[\bar{\phi}^*,\bar{\phi}]=\int_X\Bigl(\bar{\phi}^*(\bar{Z}\partial_\tau-A\nabla^2-A_{\rm z}\partial_z^2)\bar{\phi}+\bar{U}(\bar{\phi}^*\bar{\phi})\Bigr),
\end{align}
with $\nabla^2=\partial_x^2+\partial_y^2$. For concreteness we restrict to the crossover from 3D to 2D in this section, and the $k$-dependence of running couplings is understood implicitly. We write
\begin{align}
\xi = \frac{A_{\rm z}}{A},\ \eta_{\rm z} = -\frac{\dot{A}_{\rm z}}{A}.
\end{align}
The isotropic case corresponds to $\xi=1$ and $\eta=\eta_{\rm z}$. Due to $A_{\Lambda}=A_{z,\Lambda}=1$ this implies $A=A_{\rm z}$ for all $k$. This behaviour is recovered in the anisotropic parametrization for large $k$. The infrared regulator function is chosen to be
\begin{align}
 \label{AnisoReg} R_k(Q) = A \Bigl(k^2- q^2 -\xi q_z^2\Bigr)\theta\Bigl(k^2- q^2 -\xi q_z^2\Bigr)
\end{align}
with $q^2=\vec{q}^2=q_x^2+q_y^2$. The regulator is designed such that the usual replacement $p^q\to A k^2$ in the integrand is applicable.

The derivation of the flow equations for $\bar{U}$, $A$, $A_{\rm z}$, and $\bar{Z}$ is analogous to the isotropic case so that we limit the presentation to a few central statements. The $\beta$-functions for the effective potential $\bar{U}$ and frequency coefficient $\bar{Z}$ in Eqs. (\ref{fe15}) and (\ref{fe19}), respectively, need to be equipped with the modified crossover function
\begin{align}
 \nonumber  F_{\rm pbc}(\tilde{L})&=\frac{2N+1}{\tilde{L}}\Biggl[1-\frac{\eta}{4}-\frac{1}{\tilde{L}^2}\Bigl(\xi-\frac{\eta_{\rm z}}{2}\Bigr)\frac{4\pi^2}{3}N(N+1)\\
 &-\frac{\xi(2\eta_{\rm z}-\xi\eta)}{\tilde{L}^4}\frac{4\pi^4}{15}N(N+1)(-1+3N+3N^2)\Biggr]
\end{align}
instead of Eq. (\ref{fe14}). For the projection of the flow equations for $A$ and $A_{\rm z}$, respectively, we employ
\begin{align}
\dot{A} &= \frac{\partial}{\partial p^2} \dot{\bar{G}}^{-1}_{k,22}(0,\vec{p}^2=p^2,p_z=0)\Bigr|_{P=0},\\
 \dot{A}_{\rm z} &= \frac{\partial}{\partial p_z^2} \dot{\bar{G}}^{-1}_{k,22}(0,\vec{p}^2=0,p_z)\Bigr|_{P=0}.
\end{align} 
Although $q_z$ is not a continuous variable for $L<\infty$, this is the natural generalization of the appropriate infinite volume projection, see also the definition of $\dot{\bar{Z}}$ from Eq. (\ref{fe16}). For the anomalous dimensions $\eta$ and $\eta_{\rm z}$ we obtain
\begin{align}
 \nonumber \Biggl\{\begin{matrix} \eta \\ \eta_{\rm z} \end{matrix}\Biggr\} &= 8 \rho_0 \lambda^2\int_Q\frac{1}{\mbox{det}_Q^2} \Big[(2-\eta)k^2+\eta q^2+\eta_{\rm z} q_z^2\Bigr]\\
 \label{zzz1} &\times \Biggl\{ \begin{matrix} \frac{1}{d}q^2 \\ \xi^2 q_z^2 \end{matrix}\Biggr\}\delta(k^2-q^2-\xi q_z^2)\theta(k^2-q^2-\xi q_z^2).
\end{align}
For $\tilde{L}= \infty$ we can employ $\int_Q q_z^2 = \frac{1}{d}\int_Q q^2$ such that $\eta=\eta_{\rm z}$ and $\xi=1$ is a consistent solution. For $\tilde{L}<\infty$, Eqs. (\ref{zzz1}) lead to a linear set of equations for $\eta$ and $\eta_{\rm z}$, which supplements the remaining flow equations. The flow of the running coupling $\xi$ is given by
\begin{align}
 \dot{\xi} = \eta \xi -\eta_{\rm z}.
\end{align}

\begin{figure}[t!]
\centering
\begin{minipage}{0.46\textwidth}
\includegraphics[width=7.5cm]{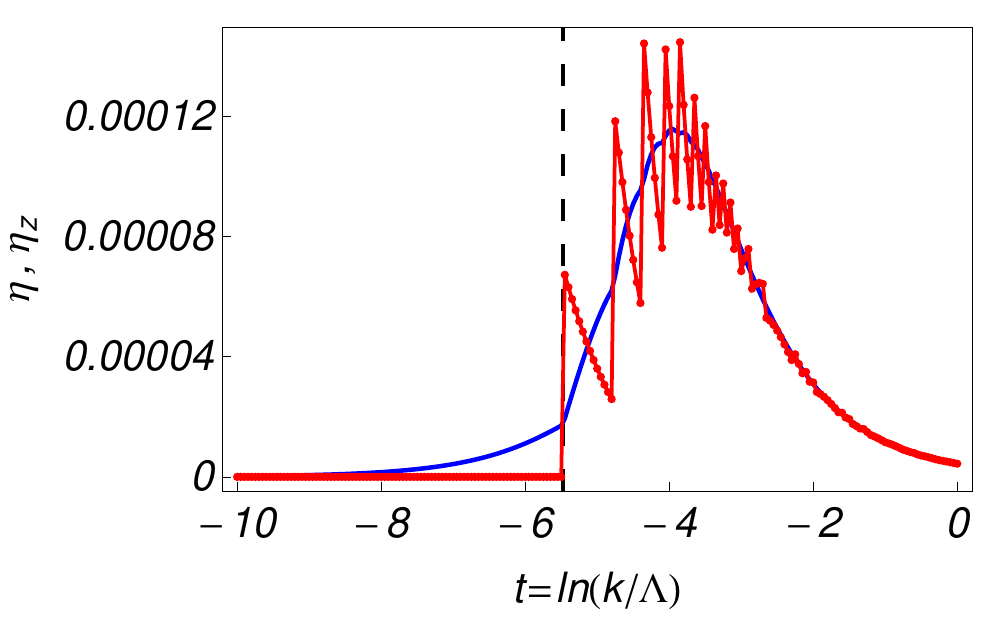}
\includegraphics[width=7cm]{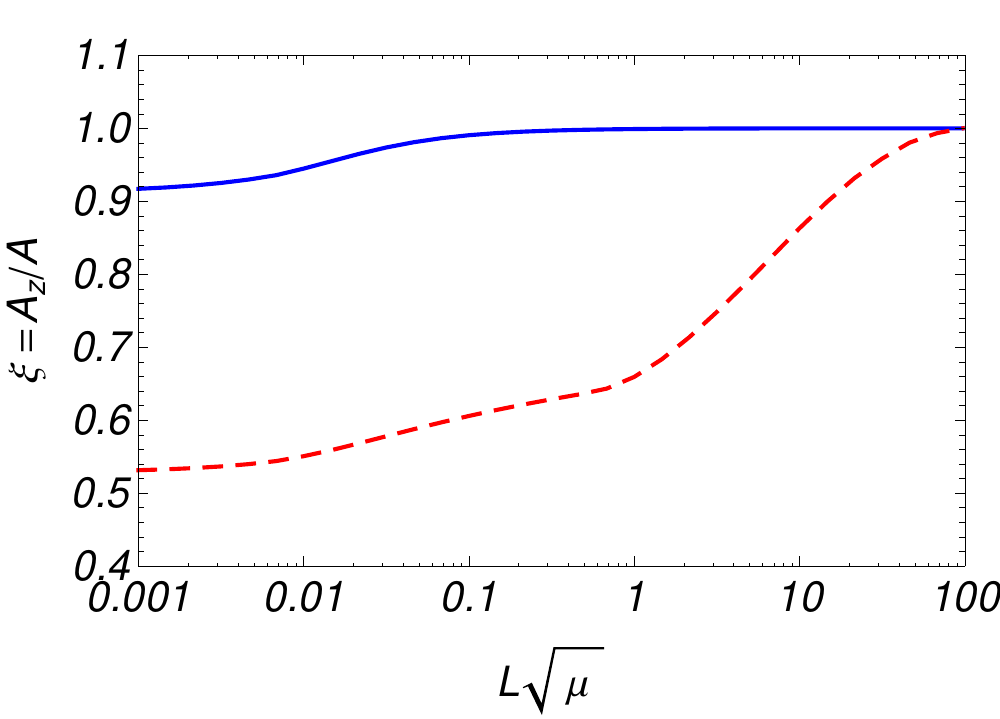}
\caption{\emph{Upper panel.} Characteristic scale dependence of $\eta$ and $\eta_{\rm z}$ within the anisotropic derivative expansion. Whereas the planar anomalous dimension $\eta$ (blue curve) remains continuous, the transverse anomalous dimension $\eta_{\rm z}$ has discontinuous jumps due to the successive integration of modes and the step function in the regulator, Eq. (\ref{AnisoReg}). The function $\eta_{\rm z}(k)$ follows the behavior of $\eta(k)$ on average for $k>\frac{2\pi}{L}$, and vanishes below $k=\frac{2\pi}{L}$ (shown as vertical dashed line). The plot is shown for $T=0$, $L\sqrt{\mu}=15$, $g_{\rm 3D}\sqrt{\mu}=0.025$. The qualitative behavior of $\eta$ and $\eta_{\rm z}$ is independent of this choice. \emph{Lower panel.} Ratio $\xi=A_{\rm z}/A$ at the lowest momentum $k_{\rm f}$ with parameters as in Fig. \ref{fig:Tc_vs_L} for zero temperature (upper curve, blue) and at $T_{\rm c}$ (lower curve, red). Since $\xi$ is of order unity  -- and not exceptionally small -- the isotropic derivative expansion is sufficient to compute observables like the critical temperature.}
\label{FigAniso}
\end{minipage}
\end{figure}

From the solution of the RG flow in the anisotropic derivative expansion we find the following features: (i) The anomalous dimension $\eta_{\rm z}$ follows the behavior of $\eta$ for $k > \frac{2\pi}{L}$ on average, though in a discontinuous manner. For $k\leq\frac{2\pi}{L}$, $\eta_{\rm z}$ vanishes, whereas $\eta$ remains nonzero. This behavior is displayed in the upper panel of Fig. \ref{FigAniso}. (ii) The value of $\xi=A_{\rm z}/A$ at $k=k_{\rm f}$ remains of order unity. It is close to $\xi=1$ at zero temperature, whereas it is in the range $\xi\simeq 0.5$-$1$ at $T_{\rm c}$, see lower panel of Fig. \ref{FigAniso}. (iii) The extracted critical temperature $T_{\rm c}$ as a function of $L$ and $g_{\rm 3D}$ is identical to the one of the isotropic derivative expansion within the resolution of Fig. \ref{fig:Tc_vs_L}. This good agreement can be understood from the fact that $\eta_{\rm z}$ closely follows the behavior of $\eta$, but $\eta$ itself is small. Hence the influence of $\eta$ on $T_{\rm c}$ is already a subleading effect, such that the difference between $\eta$ and $\eta_{\rm z}$ is only a second-order subleading effect.

The good agreement between results within the anisotropic and isotropic derivative expansions justify the use of the isotropic parametrization with $A_{\rm z}=A$ for the computation of observables in the main text.

\section{T-matrix, scattering vertex, and dimer propagator}\label{SecTmat}

In this appendix we clarify the roles played by the low-energy T-matrix, the scattering vertex, and the dimer propagator in the FRG approach to scattering of non-relativistic bosons. A key result for the analysis of the dimensional crossover is given by the translations formulas between the initial couplings $\lambda_{\Lambda}$ of the RG flow and the scattering lengths $a$.  In this section we denote the 2D and 3D scattering lengths by $a_3$ and $a_2$, respectively. We further write
\begin{align}
 \mathcal{P}_\phi(Q) = \mathcal{P}_\phi^Q =  \rmi q_0 + q^2 -\mu
\end{align}
for the inverse atom propagator with $\mu\leq 0$. The temperature is set to $T=0$ for the scattering problem.

\emph{T-matrix}. The T-matrix from Eq. (\ref{tmat}) can be defined more generally as a function of $P=(p_0,\vec{p})$ by 
\begin{align}
 \label{TP} \frac{1}{T(P)} = \frac{1}{\lambda_\Lambda^{\rm (sh)}} +\int_Q^\Lambda \frac{1}{\mathcal{P}_\phi(Q+P)\mathcal{P}_\phi(-Q)}.
\end{align}
The UV cutoff $\Lambda$ only restricts the spatial momenta. Performing the loop integration we obtain
\begin{align}
 \frac{1}{T(P)} = \begin{cases} \frac{1}{\lambda_{{\rm 3D},\Lambda}^{\rm (sh)}} +\frac{\Lambda}{4\pi^2} - \frac{1}{8\pi}\sqrt{A} & (d=3)\\ \frac{1}{\lambda_{{\rm 2D},\Lambda}^{\rm (sh)}} -\frac{1}{8\pi}\log(A/\Lambda^2)& (d=2)\end{cases}
\end{align}
with 
\begin{align}
A=\frac{\rmi p_0}{2}+\frac{p^2}{4}-\mu.
\end{align}
The analytic continuation to real frequencies (energies) $E$ is given by $\rmi p_0\to-(E+\rmi 0)$. This yields the low-energy T-matrix $T(E)$ provided we further choose $p^2=\mu=0$. To summarize, we obtain the T-matrix $T(E)$ by replacing
\begin{align}
 A \to -\frac{E+\rmi 0}{2}.
\end{align}

\emph{Scattering vertex.} By relating the running coupling constant $\lambda_k$ to the T-matrix we are able to define $a_3$ and $a_2$ within the FRG setup. The flow equation for $\lambda_k$ in vacuum ($\mu=T=0$) is found from $\dot{\lambda}_k = \dot{U}''(0)$, where
\begin{align}
 \dot{U}(\rho) = \frac{1}{2}\int_Q \dot{R}_k(Q) \frac{L_\phi^Q+L_\phi^{-Q}}{L_\phi^QL_\phi^{-Q} -(\rho U'')^2}
\end{align}
and $L_\phi^Q = \mathcal{P}_\phi(Q) + R_k(Q) +U'+\rho U''$. Hence
\begin{align}
 \nonumber \dot{\lambda}_k &= 2 \lambda_k^2 \int_Q \frac{\dot{R}_k(Q)}{(\mathcal{P}_\phi^Q+R_k^Q)^2(\mathcal{P}_\phi^{-Q}+R_k^{-Q})}\\
 &= - \lambda_k^2 \partial_t \int_Q \frac{1}{(\mathcal{P}_\phi^Q+R_k^Q)(\mathcal{P}_\phi^{-Q}+R_k^{-Q})}.
\end{align}
We conclude that
\begin{align}
\partial_t \frac{1}{\lambda_k} = \partial_t  \int_Q \frac{1}{(\mathcal{P}_\phi^Q+R_k^Q)(\mathcal{P}_\phi^{-Q}+R_k^{-Q})},
\end{align}
which is readily solved by
\begin{align}
 \nonumber \frac{1}{\lambda_{k}} &= \frac{1}{\lambda_{\Lambda}} + \int_Q \Bigl(\frac{1}{(\mathcal{P}_\phi^Q+R_k^Q)(\mathcal{P}_\phi^{-Q}+R_k^{-Q})}\\
 \nonumber & -\frac{1}{(\mathcal{P}_\phi^Q+R_\Lambda^Q)(\mathcal{P}_\phi^{-Q}+R_\Lambda^{-Q})}\Bigr)\\
 \nonumber &=\frac{1}{\lambda_{\Lambda}} +\frac{1}{2} \int_{\vec{q}} \Bigl(\frac{1}{q^2+R_k(q)} -\frac{1}{q^2+R_\Lambda(q)}\Bigr)\\
 \label{t1} &\stackrel{(\ref{t2})}{=}\frac{1}{\lambda_{\Lambda}} +\frac{1}{2} \int_{\vec{q}}^\Lambda \frac{1}{q^2+R_k(q)} -\frac{1}{2}\int_{\vec{q}}^\Lambda\frac{1}{\Lambda^2}.
\end{align}
In the last line we employed the particular form of the Litim-type regulator given by
\begin{align}
 \label{t2} R_k(q)=(k^2-q^2)\theta(k^2-q^2).
\end{align}
For the last term in Eq. (\ref{t1}) we find
\begin{align}
 \frac{1}{2}\int_{\vec{q}}^\Lambda\frac{1}{\Lambda^2} = \begin{cases} \frac{\Lambda}{12\pi^2}& (d=3),\\ \frac{1}{8\pi} &(d=2).\end{cases}
\end{align}
We generalize this to a $P$-dependent scattering vertex by defining
\begin{align}
\partial_t \frac{1}{\lambda_k(P)} := \partial_t  \int_Q \frac{1}{(\mathcal{P}_\phi^{Q+P}+R_k^{Q+P})(\mathcal{P}_\phi^{-Q}+R_k^{-Q})}.
\end{align}
For nonzero $P$ we can take the limit $k\to0$ and find
\begin{align}
 \nonumber \frac{1}{\lambda_{k=0}(P)} &= \frac{1}{\lambda_{\Lambda}} + \int_Q^\Lambda \frac{1}{\mathcal{P}_\phi^{Q+P}\mathcal{P}_\phi^{-Q}}-\frac{1}{2}\int_{\vec{q}}^\Lambda\frac{1}{\Lambda^2}
\end{align}
due to $R_{k=0}=0$. Comparing this to Eq. (\ref{TP}) for the T-matrix $T(P)$ we conclude that 
\begin{align}
 T(P)&=\lambda_{k=0}(P).
\end{align}
For the renormalization of the coupling constants we have
\begin{align}
  \frac{1}{\lambda_\Lambda^{\rm (sh)}}&=\frac{1}{\lambda_{\Lambda}}-\frac{1}{2}\int_{\vec{q}}^\Lambda \frac{1}{\Lambda^2}.
\end{align}
The additional term on the right hand side results from the particular form of the regulator (\ref{t2}) employed in our FRG approach. As a consequence, the relationship between $\lambda_{\Lambda}$ and the scattering length $a$ in the FRG approach is given by
\begin{align}
 \label{t3a}\frac{1}{\lambda_{\Lambda}^{\rm 3D}} &= -\frac{\Lambda}{6\pi^2} + \frac{1}{8\pi a_3},\\
 \label{t3b}\frac{1}{\lambda_{\Lambda}^{\rm 2D}} &= -\frac{1}{8\pi}\log(\Lambda^2 a_2^2) +\frac{1}{8\pi}.
\end{align}

\emph{Dimer propagator.} The scattering vertex $\lambda_k(P)$  has an intuitive interpretation in terms of the inverse dimer propagator
\begin{align}
 \mathcal{P}_{{\rm dim}}(P) := -\frac{1}{\lambda_{k=0}(P)} =-\frac{1}{T(P)},
\end{align}
which describes interactions between atoms by means of a propagating dimer boson with frequency and momentum $P$. Within the sharp momentum cutoff scheme we have
\begin{align}
  \mathcal{P}_{{\rm dim}}(P)  = - \frac{1}{\lambda_\Lambda^{\rm (sh)}} - \int_Q^\Lambda \frac{1}{\mathcal{P}_\phi(Q+P)\mathcal{P}_\phi(-Q)}.
\end{align}
Due to Galilean invariance, the $P$-dependence of the dimer propagator is solely in terms of
\begin{align}
 \label{Aeq} A = \frac{1}{2}\Bigl(  \rmi p_0 + \frac{p^2}{2}-2\mu \Bigr),
\end{align}
which reflects the dimer mass and dimer chemical potential being $M_{\rm dim}=2M$ and $\mu_{\rm dim}=2\mu$, respectively.

\end{appendix}


\bibliographystyle{apsrev4-1}
\bibliography{references_dimcrossoverBose} 

\begin{thebibliography}{80}%
\makeatletter
\providecommand \@ifxundefined [1]{%
 \@ifx{#1\undefined}
}%
\providecommand \@ifnum [1]{%
 \ifnum #1\expandafter \@firstoftwo
 \else \expandafter \@secondoftwo
 \fi
}%
\providecommand \@ifx [1]{%
 \ifx #1\expandafter \@firstoftwo
 \else \expandafter \@secondoftwo
 \fi
}%
\providecommand \natexlab [1]{#1}%
\providecommand \enquote  [1]{``#1''}%
\providecommand \bibnamefont  [1]{#1}%
\providecommand \bibfnamefont [1]{#1}%
\providecommand \citenamefont [1]{#1}%
\providecommand \href@noop [0]{\@secondoftwo}%
\providecommand \href [0]{\begingroup \@sanitize@url \@href}%
\providecommand \@href[1]{\@@startlink{#1}\@@href}%
\providecommand \@@href[1]{\endgroup#1\@@endlink}%
\providecommand \@sanitize@url [0]{\catcode `\\12\catcode `\$12\catcode
  `\&12\catcode `\#12\catcode `\^12\catcode `\_12\catcode `\%12\relax}%
\providecommand \@@startlink[1]{}%
\providecommand \@@endlink[0]{}%
\providecommand \url  [0]{\begingroup\@sanitize@url \@url }%
\providecommand \@url [1]{\endgroup\@href {#1}{\urlprefix }}%
\providecommand \urlprefix  [0]{URL }%
\providecommand \Eprint [0]{\href }%
\providecommand \doibase [0]{http://dx.doi.org/}%
\providecommand \selectlanguage [0]{\@gobble}%
\providecommand \bibinfo  [0]{\@secondoftwo}%
\providecommand \bibfield  [0]{\@secondoftwo}%
\providecommand \translation [1]{[#1]}%
\providecommand \BibitemOpen [0]{}%
\providecommand \bibitemStop [0]{}%
\providecommand \bibitemNoStop [0]{.\EOS\space}%
\providecommand \EOS [0]{\spacefactor3000\relax}%
\providecommand \BibitemShut  [1]{\csname bibitem#1\endcsname}%
\let\auto@bib@innerbib\@empty
\bibitem [{\citenamefont {Bloch}\ \emph {et~al.}(2008)\citenamefont {Bloch},
  \citenamefont {Dalibard},\ and\ \citenamefont {Zwerger}}]{RevModPhys.80.885}%
  \BibitemOpen
  \bibfield  {author} {\bibinfo {author} {\bibfnamefont {I.}~\bibnamefont
  {Bloch}}, \bibinfo {author} {\bibfnamefont {J.}~\bibnamefont {Dalibard}}, \
  and\ \bibinfo {author} {\bibfnamefont {W.}~\bibnamefont {Zwerger}},\ }\href
  {\doibase 10.1103/RevModPhys.80.885} {\bibfield  {journal} {\bibinfo
  {journal} {Rev. Mod. Phys.}\ }\textbf {\bibinfo {volume} {80}},\ \bibinfo
  {pages} {885} (\bibinfo {year} {2008})}\BibitemShut {NoStop}%
\bibitem [{\citenamefont {Lewenstein}\ \emph {et~al.}(2006)\citenamefont
  {Lewenstein}, \citenamefont {Sanpera}, \citenamefont {Ahufinger},
  \citenamefont {Damski}, \citenamefont {De},\ and\ \citenamefont
  {Sen}}]{lewenstein-rmp-56-135}%
  \BibitemOpen
  \bibfield  {author} {\bibinfo {author} {\bibfnamefont {M.}~\bibnamefont
  {Lewenstein}}, \bibinfo {author} {\bibfnamefont {A.}~\bibnamefont {Sanpera}},
  \bibinfo {author} {\bibfnamefont {V.}~\bibnamefont {Ahufinger}}, \bibinfo
  {author} {\bibfnamefont {B.}~\bibnamefont {Damski}}, \bibinfo {author}
  {\bibfnamefont {A.~S.}\ \bibnamefont {De}}, \ and\ \bibinfo {author}
  {\bibfnamefont {U.}~\bibnamefont {Sen}},\ }\href@noop {} {\bibfield
  {journal} {\bibinfo  {journal} {Advances in Physics}\ }\textbf {\bibinfo
  {volume} {56}},\ \bibinfo {pages} {135} (\bibinfo {year} {2006})}\BibitemShut
  {NoStop}%
\bibitem [{\citenamefont {Hadzibabic}\ \emph {et~al.}(2006)\citenamefont
  {Hadzibabic}, \citenamefont {Kr\"{u}ger}, \citenamefont {Cheneau},
  \citenamefont {Battelier},\ and\ \citenamefont {Dalibard}}]{Hadzibabic2006}%
  \BibitemOpen
  \bibfield  {author} {\bibinfo {author} {\bibfnamefont {Z.}~\bibnamefont
  {Hadzibabic}}, \bibinfo {author} {\bibfnamefont {P.}~\bibnamefont
  {Kr\"{u}ger}}, \bibinfo {author} {\bibfnamefont {M.}~\bibnamefont {Cheneau}},
  \bibinfo {author} {\bibfnamefont {B.}~\bibnamefont {Battelier}}, \ and\
  \bibinfo {author} {\bibfnamefont {J.}~\bibnamefont {Dalibard}},\ }\href
  {\doibase 10.1038/nature04851} {\bibfield  {journal} {\bibinfo  {journal}
  {Nature}\ }\textbf {\bibinfo {volume} {441}},\ \bibinfo {pages} {1118}
  (\bibinfo {year} {2006})}\BibitemShut {NoStop}%
\bibitem [{\citenamefont {Clad\'e}\ \emph {et~al.}(2009)\citenamefont
  {Clad\'e}, \citenamefont {Ryu}, \citenamefont {Ramanathan}, \citenamefont
  {Helmerson},\ and\ \citenamefont {Phillips}}]{Clade2009}%
  \BibitemOpen
  \bibfield  {author} {\bibinfo {author} {\bibfnamefont {P.}~\bibnamefont
  {Clad\'e}}, \bibinfo {author} {\bibfnamefont {C.}~\bibnamefont {Ryu}},
  \bibinfo {author} {\bibfnamefont {A.}~\bibnamefont {Ramanathan}}, \bibinfo
  {author} {\bibfnamefont {K.}~\bibnamefont {Helmerson}}, \ and\ \bibinfo
  {author} {\bibfnamefont {W.~D.}\ \bibnamefont {Phillips}},\ }\href {\doibase
  10.1103/PhysRevLett.102.170401} {\bibfield  {journal} {\bibinfo  {journal}
  {Phys. Rev. Lett.}\ }\textbf {\bibinfo {volume} {102}},\ \bibinfo {pages}
  {170401} (\bibinfo {year} {2009})}\BibitemShut {NoStop}%
\bibitem [{\citenamefont {Tung}\ \emph {et~al.}(2010)\citenamefont {Tung},
  \citenamefont {Lamporesi}, \citenamefont {Lobser}, \citenamefont {Xia},\ and\
  \citenamefont {Cornell}}]{Tung2010}%
  \BibitemOpen
  \bibfield  {author} {\bibinfo {author} {\bibfnamefont {S.}~\bibnamefont
  {Tung}}, \bibinfo {author} {\bibfnamefont {G.}~\bibnamefont {Lamporesi}},
  \bibinfo {author} {\bibfnamefont {D.}~\bibnamefont {Lobser}}, \bibinfo
  {author} {\bibfnamefont {L.}~\bibnamefont {Xia}}, \ and\ \bibinfo {author}
  {\bibfnamefont {E.~A.}\ \bibnamefont {Cornell}},\ }\href {\doibase
  10.1103/PhysRevLett.105.230408} {\bibfield  {journal} {\bibinfo  {journal}
  {Phys. Rev. Lett.}\ }\textbf {\bibinfo {volume} {105}},\ \bibinfo {pages}
  {230408} (\bibinfo {year} {2010})}\BibitemShut {NoStop}%
\bibitem [{\citenamefont {Plisson}\ \emph {et~al.}(2011)\citenamefont
  {Plisson}, \citenamefont {Allard}, \citenamefont {Holzmann}, \citenamefont
  {Salomon}, \citenamefont {Aspect}, \citenamefont {Bouyer},\ and\
  \citenamefont {Bourdel}}]{Plisson2011}%
  \BibitemOpen
  \bibfield  {author} {\bibinfo {author} {\bibfnamefont {T.}~\bibnamefont
  {Plisson}}, \bibinfo {author} {\bibfnamefont {B.}~\bibnamefont {Allard}},
  \bibinfo {author} {\bibfnamefont {M.}~\bibnamefont {Holzmann}}, \bibinfo
  {author} {\bibfnamefont {G.}~\bibnamefont {Salomon}}, \bibinfo {author}
  {\bibfnamefont {A.}~\bibnamefont {Aspect}}, \bibinfo {author} {\bibfnamefont
  {P.}~\bibnamefont {Bouyer}}, \ and\ \bibinfo {author} {\bibfnamefont
  {T.}~\bibnamefont {Bourdel}},\ }\href {\doibase 10.1103/PhysRevA.84.061606}
  {\bibfield  {journal} {\bibinfo  {journal} {Phys. Rev. A}\ }\textbf {\bibinfo
  {volume} {84}},\ \bibinfo {pages} {061606} (\bibinfo {year}
  {2011})}\BibitemShut {NoStop}%
\bibitem [{\citenamefont {Desbuquois}\ \emph {et~al.}(2012)\citenamefont
  {Desbuquois}, \citenamefont {Chomaz}, \citenamefont {Yefsah}, \citenamefont
  {L\'{e}onard}, \citenamefont {Beugnon}, \citenamefont {Weitenberg},\ and\
  \citenamefont {Dalibard}}]{Desbuquois2012}%
  \BibitemOpen
  \bibfield  {author} {\bibinfo {author} {\bibfnamefont {R.}~\bibnamefont
  {Desbuquois}}, \bibinfo {author} {\bibfnamefont {L.}~\bibnamefont {Chomaz}},
  \bibinfo {author} {\bibfnamefont {T.}~\bibnamefont {Yefsah}}, \bibinfo
  {author} {\bibfnamefont {J.}~\bibnamefont {L\'{e}onard}}, \bibinfo {author}
  {\bibfnamefont {J.}~\bibnamefont {Beugnon}}, \bibinfo {author} {\bibfnamefont
  {C.}~\bibnamefont {Weitenberg}}, \ and\ \bibinfo {author} {\bibfnamefont
  {J.}~\bibnamefont {Dalibard}},\ }\href {\doibase 10.1038/nphys2378}
  {\bibfield  {journal} {\bibinfo  {journal} {Nat. Phys.}\ }\textbf {\bibinfo
  {volume} {8}},\ \bibinfo {pages} {645} (\bibinfo {year} {2012})}\BibitemShut
  {NoStop}%
\bibitem [{\citenamefont {Fletcher}\ \emph {et~al.}(2015)\citenamefont
  {Fletcher}, \citenamefont {Robert-de Saint-Vincent}, \citenamefont {Man},
  \citenamefont {Navon}, \citenamefont {Smith}, \citenamefont {Viebahn},\ and\
  \citenamefont {Hadzibabic}}]{PhysRevLett.114.255302}%
  \BibitemOpen
  \bibfield  {author} {\bibinfo {author} {\bibfnamefont {R.~J.}\ \bibnamefont
  {Fletcher}}, \bibinfo {author} {\bibfnamefont {M.}~\bibnamefont {Robert-de
  Saint-Vincent}}, \bibinfo {author} {\bibfnamefont {J.}~\bibnamefont {Man}},
  \bibinfo {author} {\bibfnamefont {N.}~\bibnamefont {Navon}}, \bibinfo
  {author} {\bibfnamefont {R.~P.}\ \bibnamefont {Smith}}, \bibinfo {author}
  {\bibfnamefont {K.~G.~H.}\ \bibnamefont {Viebahn}}, \ and\ \bibinfo {author}
  {\bibfnamefont {Z.}~\bibnamefont {Hadzibabic}},\ }\href {\doibase
  10.1103/PhysRevLett.114.255302} {\bibfield  {journal} {\bibinfo  {journal}
  {Phys. Rev. Lett.}\ }\textbf {\bibinfo {volume} {114}},\ \bibinfo {pages}
  {255302} (\bibinfo {year} {2015})}\BibitemShut {NoStop}%
\bibitem [{\citenamefont {Ries}\ \emph {et~al.}(2015)\citenamefont {Ries},
  \citenamefont {Wenz}, \citenamefont {Z\"urn}, \citenamefont {Bayha},
  \citenamefont {Boettcher}, \citenamefont {Kedar}, \citenamefont {Murthy},
  \citenamefont {Neidig}, \citenamefont {Lompe},\ and\ \citenamefont
  {Jochim}}]{PhysRevLett.114.230401}%
  \BibitemOpen
  \bibfield  {author} {\bibinfo {author} {\bibfnamefont {M.~G.}\ \bibnamefont
  {Ries}}, \bibinfo {author} {\bibfnamefont {A.~N.}\ \bibnamefont {Wenz}},
  \bibinfo {author} {\bibfnamefont {G.}~\bibnamefont {Z\"urn}}, \bibinfo
  {author} {\bibfnamefont {L.}~\bibnamefont {Bayha}}, \bibinfo {author}
  {\bibfnamefont {I.}~\bibnamefont {Boettcher}}, \bibinfo {author}
  {\bibfnamefont {D.}~\bibnamefont {Kedar}}, \bibinfo {author} {\bibfnamefont
  {P.~A.}\ \bibnamefont {Murthy}}, \bibinfo {author} {\bibfnamefont
  {M.}~\bibnamefont {Neidig}}, \bibinfo {author} {\bibfnamefont
  {T.}~\bibnamefont {Lompe}}, \ and\ \bibinfo {author} {\bibfnamefont
  {S.}~\bibnamefont {Jochim}},\ }\href {\doibase
  10.1103/PhysRevLett.114.230401} {\bibfield  {journal} {\bibinfo  {journal}
  {Phys. Rev. Lett.}\ }\textbf {\bibinfo {volume} {114}},\ \bibinfo {pages}
  {230401} (\bibinfo {year} {2015})}\BibitemShut {NoStop}%
\bibitem [{\citenamefont {Murthy}\ \emph {et~al.}(2015)\citenamefont {Murthy},
  \citenamefont {Boettcher}, \citenamefont {Bayha}, \citenamefont {Holzmann},
  \citenamefont {Kedar}, \citenamefont {Neidig}, \citenamefont {Ries},
  \citenamefont {Wenz}, \citenamefont {Z\"urn},\ and\ \citenamefont
  {Jochim}}]{PhysRevLett.115.010401}%
  \BibitemOpen
  \bibfield  {author} {\bibinfo {author} {\bibfnamefont {P.~A.}\ \bibnamefont
  {Murthy}}, \bibinfo {author} {\bibfnamefont {I.}~\bibnamefont {Boettcher}},
  \bibinfo {author} {\bibfnamefont {L.}~\bibnamefont {Bayha}}, \bibinfo
  {author} {\bibfnamefont {M.}~\bibnamefont {Holzmann}}, \bibinfo {author}
  {\bibfnamefont {D.}~\bibnamefont {Kedar}}, \bibinfo {author} {\bibfnamefont
  {M.}~\bibnamefont {Neidig}}, \bibinfo {author} {\bibfnamefont {M.~G.}\
  \bibnamefont {Ries}}, \bibinfo {author} {\bibfnamefont {A.~N.}\ \bibnamefont
  {Wenz}}, \bibinfo {author} {\bibfnamefont {G.}~\bibnamefont {Z\"urn}}, \ and\
  \bibinfo {author} {\bibfnamefont {S.}~\bibnamefont {Jochim}},\ }\href
  {\doibase 10.1103/PhysRevLett.115.010401} {\bibfield  {journal} {\bibinfo
  {journal} {Phys. Rev. Lett.}\ }\textbf {\bibinfo {volume} {115}},\ \bibinfo
  {pages} {010401} (\bibinfo {year} {2015})}\BibitemShut {NoStop}%
\bibitem [{\citenamefont {Wetterich}(1991)}]{Wetterich:1989xg}%
  \BibitemOpen
  \bibfield  {author} {\bibinfo {author} {\bibfnamefont {C.}~\bibnamefont
  {Wetterich}},\ }\href {\doibase 10.1016/0550-3213(91)90099-J} {\bibfield
  {journal} {\bibinfo  {journal} {Nucl.Phys.}\ }\textbf {\bibinfo {volume}
  {B352}},\ \bibinfo {pages} {529} (\bibinfo {year} {1991})}\BibitemShut
  {NoStop}%
\bibitem [{\citenamefont {Wetterich}(1993)}]{Wetterich1993}%
  \BibitemOpen
  \bibfield  {author} {\bibinfo {author} {\bibfnamefont {C.}~\bibnamefont
  {Wetterich}},\ }\href {\doibase 10.1016/0370-2693(93)90726-X} {\bibfield
  {journal} {\bibinfo  {journal} {Physics Letters B}\ }\textbf {\bibinfo
  {volume} {301}},\ \bibinfo {pages} {90} (\bibinfo {year} {1993})}\BibitemShut
  {NoStop}%
\bibitem [{\citenamefont {Berges}\ \emph {et~al.}(2002)\citenamefont {Berges},
  \citenamefont {Tetradis},\ and\ \citenamefont {Wetterich}}]{Berges:2000ew}%
  \BibitemOpen
  \bibfield  {author} {\bibinfo {author} {\bibfnamefont {J.}~\bibnamefont
  {Berges}}, \bibinfo {author} {\bibfnamefont {N.}~\bibnamefont {Tetradis}}, \
  and\ \bibinfo {author} {\bibfnamefont {C.}~\bibnamefont {Wetterich}},\ }\href
  {\doibase 10.1016/S0370-1573(01)00098-9} {\bibfield  {journal} {\bibinfo
  {journal} {Phys.Rept.}\ }\textbf {\bibinfo {volume} {363}},\ \bibinfo {pages}
  {223} (\bibinfo {year} {2002})}\BibitemShut {NoStop}%
\bibitem [{\citenamefont {Pawlowski}(2007)}]{Pawlowski20072831}%
  \BibitemOpen
  \bibfield  {author} {\bibinfo {author} {\bibfnamefont {J.~M.}\ \bibnamefont
  {Pawlowski}},\ }\href {\doibase 10.1016/j.aop.2007.01.007} {\bibfield
  {journal} {\bibinfo  {journal} {Annals of Physics}\ }\textbf {\bibinfo
  {volume} {322}},\ \bibinfo {pages} {2831 } (\bibinfo {year}
  {2007})}\BibitemShut {NoStop}%
\bibitem [{\citenamefont {Gies}(2012)}]{Gies:2006wv}%
  \BibitemOpen
  \bibfield  {author} {\bibinfo {author} {\bibfnamefont {H.}~\bibnamefont
  {Gies}},\ }\href {\doibase 10.1007/978-3-642-27320-9_6} {\bibfield  {journal}
  {\bibinfo  {journal} {Lect.Notes Phys.}\ }\textbf {\bibinfo {volume} {852}},\
  \bibinfo {pages} {287} (\bibinfo {year} {2012})}\BibitemShut {NoStop}%
\bibitem [{\citenamefont {Schaefer}\ and\ \citenamefont
  {Wambach}(2008)}]{Schaefer:2006sr}%
  \BibitemOpen
  \bibfield  {author} {\bibinfo {author} {\bibfnamefont {B.-J.}\ \bibnamefont
  {Schaefer}}\ and\ \bibinfo {author} {\bibfnamefont {J.}~\bibnamefont
  {Wambach}},\ }\href {\doibase 10.1134/S1063779608070083} {\bibfield
  {journal} {\bibinfo  {journal} {Phys.Part.Nucl.}\ }\textbf {\bibinfo {volume}
  {39}},\ \bibinfo {pages} {1025} (\bibinfo {year} {2008})}\BibitemShut
  {NoStop}%
\bibitem [{\citenamefont {Delamotte}(2012)}]{Delamotte:2007pf}%
  \BibitemOpen
  \bibfield  {author} {\bibinfo {author} {\bibfnamefont {B.}~\bibnamefont
  {Delamotte}},\ }\href {\doibase 10.1007/978-3-642-27320-9_2} {\bibfield
  {journal} {\bibinfo  {journal} {Lect.Notes Phys.}\ }\textbf {\bibinfo
  {volume} {852}},\ \bibinfo {pages} {49} (\bibinfo {year} {2012})}\BibitemShut
  {NoStop}%
\bibitem [{\citenamefont {Kopietz}\ \emph {et~al.}(2010)\citenamefont
  {Kopietz}, \citenamefont {Bartosch},\ and\ \citenamefont
  {Sch\"{u}tz}}]{Kopietz2010}%
  \BibitemOpen
  \bibfield  {author} {\bibinfo {author} {\bibfnamefont {P.}~\bibnamefont
  {Kopietz}}, \bibinfo {author} {\bibfnamefont {L.}~\bibnamefont {Bartosch}}, \
  and\ \bibinfo {author} {\bibfnamefont {F.}~\bibnamefont {Sch\"{u}tz}},\
  }\href@noop {} {\emph {\bibinfo {title} {{Introduction to the Functional
  Renormalization Group}}}}\ (\bibinfo  {publisher} {Springer, Berlin},\
  \bibinfo {year} {2010})\BibitemShut {NoStop}%
\bibitem [{\citenamefont {Metzner}\ \emph {et~al.}(2012)\citenamefont
  {Metzner}, \citenamefont {Salmhofer}, \citenamefont {Honerkamp},
  \citenamefont {Meden},\ and\ \citenamefont {Schonhammer}}]{Metzner:2011cw}%
  \BibitemOpen
  \bibfield  {author} {\bibinfo {author} {\bibfnamefont {W.}~\bibnamefont
  {Metzner}}, \bibinfo {author} {\bibfnamefont {M.}~\bibnamefont {Salmhofer}},
  \bibinfo {author} {\bibfnamefont {C.}~\bibnamefont {Honerkamp}}, \bibinfo
  {author} {\bibfnamefont {V.}~\bibnamefont {Meden}}, \ and\ \bibinfo {author}
  {\bibfnamefont {K.}~\bibnamefont {Schonhammer}},\ }\href@noop {} {\bibfield
  {journal} {\bibinfo  {journal} {Rev.Mod.Phys.}\ }\textbf {\bibinfo {volume}
  {84}},\ \bibinfo {pages} {299} (\bibinfo {year} {2012})}\BibitemShut
  {NoStop}%
\bibitem [{\citenamefont {Braun}(2012)}]{Braun:2011pp}%
  \BibitemOpen
  \bibfield  {author} {\bibinfo {author} {\bibfnamefont {J.}~\bibnamefont
  {Braun}},\ }\href {\doibase 10.1088/0954-3899/39/3/033001} {\bibfield
  {journal} {\bibinfo  {journal} {J.Phys.}\ }\textbf {\bibinfo {volume}
  {G39}},\ \bibinfo {pages} {033001} (\bibinfo {year} {2012})}\BibitemShut
  {NoStop}%
\bibitem [{\citenamefont {Boettcher}\ \emph {et~al.}(2012)\citenamefont
  {Boettcher}, \citenamefont {Pawlowski},\ and\ \citenamefont
  {Diehl}}]{Boettcher:2012cm}%
  \BibitemOpen
  \bibfield  {author} {\bibinfo {author} {\bibfnamefont {I.}~\bibnamefont
  {Boettcher}}, \bibinfo {author} {\bibfnamefont {J.~M.}\ \bibnamefont
  {Pawlowski}}, \ and\ \bibinfo {author} {\bibfnamefont {S.}~\bibnamefont
  {Diehl}},\ }\href {\doibase 10.1016/j.nuclphysbps.2012.06.004} {\bibfield
  {journal} {\bibinfo  {journal} {Nucl.Phys.Proc.Suppl.}\ }\textbf {\bibinfo
  {volume} {228}},\ \bibinfo {pages} {63} (\bibinfo {year} {2012})}\BibitemShut
  {NoStop}%
\bibitem [{\citenamefont {Pawlowski}\ \emph {et~al.}(2015)\citenamefont
  {Pawlowski}, \citenamefont {Scherer}, \citenamefont {Schmidt},\ and\
  \citenamefont {Wetzel}}]{Pawlowski:2015mlf}%
  \BibitemOpen
  \bibfield  {author} {\bibinfo {author} {\bibfnamefont {J.~M.}\ \bibnamefont
  {Pawlowski}}, \bibinfo {author} {\bibfnamefont {M.~M.}\ \bibnamefont
  {Scherer}}, \bibinfo {author} {\bibfnamefont {R.}~\bibnamefont {Schmidt}}, \
  and\ \bibinfo {author} {\bibfnamefont {S.~J.}\ \bibnamefont {Wetzel}},\
  }\href@noop {} {\  (\bibinfo {year} {2015})},\ \Eprint
  {http://arxiv.org/abs/1512.03598} {arXiv:1512.03598} \BibitemShut {NoStop}%
\bibitem [{\citenamefont {Wetterich}(2008)}]{PhysRevB.77.064504}%
  \BibitemOpen
  \bibfield  {author} {\bibinfo {author} {\bibfnamefont {C.}~\bibnamefont
  {Wetterich}},\ }\href {\doibase 10.1103/PhysRevB.77.064504} {\bibfield
  {journal} {\bibinfo  {journal} {Phys. Rev. B}\ }\textbf {\bibinfo {volume}
  {77}},\ \bibinfo {pages} {064504} (\bibinfo {year} {2008})}\BibitemShut
  {NoStop}%
\bibitem [{\citenamefont {Floerchinger}\ and\ \citenamefont
  {Wetterich}(2008)}]{PhysRevA.77.053603}%
  \BibitemOpen
  \bibfield  {author} {\bibinfo {author} {\bibfnamefont {S.}~\bibnamefont
  {Floerchinger}}\ and\ \bibinfo {author} {\bibfnamefont {C.}~\bibnamefont
  {Wetterich}},\ }\href {\doibase 10.1103/PhysRevA.77.053603} {\bibfield
  {journal} {\bibinfo  {journal} {Phys. Rev. A}\ }\textbf {\bibinfo {volume}
  {77}},\ \bibinfo {pages} {053603} (\bibinfo {year} {2008})}\BibitemShut
  {NoStop}%
\bibitem [{\citenamefont {Floerchinger}\ and\ \citenamefont
  {Wetterich}(2009{\natexlab{a}})}]{PhysRevA.79.013601}%
  \BibitemOpen
  \bibfield  {author} {\bibinfo {author} {\bibfnamefont {S.}~\bibnamefont
  {Floerchinger}}\ and\ \bibinfo {author} {\bibfnamefont {C.}~\bibnamefont
  {Wetterich}},\ }\href {\doibase 10.1103/PhysRevA.79.013601} {\bibfield
  {journal} {\bibinfo  {journal} {Phys. Rev. A}\ }\textbf {\bibinfo {volume}
  {79}},\ \bibinfo {pages} {013601} (\bibinfo {year}
  {2009}{\natexlab{a}})}\BibitemShut {NoStop}%
\bibitem [{\citenamefont {Mermin}\ and\ \citenamefont
  {Wagner}(1966)}]{Mermin1966}%
  \BibitemOpen
  \bibfield  {author} {\bibinfo {author} {\bibfnamefont {N.~D.}\ \bibnamefont
  {Mermin}}\ and\ \bibinfo {author} {\bibfnamefont {H.}~\bibnamefont
  {Wagner}},\ }\href {\doibase 10.1103/PhysRevLett.17.1133} {\bibfield
  {journal} {\bibinfo  {journal} {Phys. Rev. Lett.}\ }\textbf {\bibinfo
  {volume} {17}},\ \bibinfo {pages} {1133} (\bibinfo {year}
  {1966})}\BibitemShut {NoStop}%
\bibitem [{\citenamefont {Hohenberg}(1967)}]{Hohenberg1967}%
  \BibitemOpen
  \bibfield  {author} {\bibinfo {author} {\bibfnamefont {P.~C.}\ \bibnamefont
  {Hohenberg}},\ }\href {\doibase 10.1103/PhysRev.158.383} {\bibfield
  {journal} {\bibinfo  {journal} {Phys. Rev.}\ }\textbf {\bibinfo {volume}
  {158}},\ \bibinfo {pages} {383} (\bibinfo {year} {1967})}\BibitemShut
  {NoStop}%
\bibitem [{\citenamefont {Floerchinger}\ and\ \citenamefont
  {Wetterich}(2009{\natexlab{b}})}]{PhysRevA.79.063602}%
  \BibitemOpen
  \bibfield  {author} {\bibinfo {author} {\bibfnamefont {S.}~\bibnamefont
  {Floerchinger}}\ and\ \bibinfo {author} {\bibfnamefont {C.}~\bibnamefont
  {Wetterich}},\ }\href {\doibase 10.1103/PhysRevA.79.063602} {\bibfield
  {journal} {\bibinfo  {journal} {Phys. Rev. A}\ }\textbf {\bibinfo {volume}
  {79}},\ \bibinfo {pages} {063602} (\bibinfo {year}
  {2009}{\natexlab{b}})}\BibitemShut {NoStop}%
\bibitem [{\citenamefont {Ran\ifmmode~\mbox{\c{c}}\else \c{c}\fi{}on}\ and\
  \citenamefont {Dupuis}(2012)}]{PhysRevA.85.063607}%
  \BibitemOpen
  \bibfield  {author} {\bibinfo {author} {\bibfnamefont {A.}~\bibnamefont
  {Ran\ifmmode~\mbox{\c{c}}\else \c{c}\fi{}on}}\ and\ \bibinfo {author}
  {\bibfnamefont {N.}~\bibnamefont {Dupuis}},\ }\href {\doibase
  10.1103/PhysRevA.85.063607} {\bibfield  {journal} {\bibinfo  {journal} {Phys.
  Rev. A}\ }\textbf {\bibinfo {volume} {85}},\ \bibinfo {pages} {063607}
  (\bibinfo {year} {2012})}\BibitemShut {NoStop}%
\bibitem [{\citenamefont {Tetradis}\ and\ \citenamefont
  {Wetterich}(1994{\natexlab{a}})}]{Tetradis1994541}%
  \BibitemOpen
  \bibfield  {author} {\bibinfo {author} {\bibfnamefont {N.}~\bibnamefont
  {Tetradis}}\ and\ \bibinfo {author} {\bibfnamefont {C.}~\bibnamefont
  {Wetterich}},\ }\href {\doibase /10.1016/0550-3213(94)90446-4} {\bibfield
  {journal} {\bibinfo  {journal} {Nuclear Physics B}\ }\textbf {\bibinfo
  {volume} {422}},\ \bibinfo {pages} {541 } (\bibinfo {year}
  {1994}{\natexlab{a}})}\BibitemShut {NoStop}%
\bibitem [{\citenamefont {Prokof'ev}\ \emph {et~al.}(2001)\citenamefont
  {Prokof'ev}, \citenamefont {Ruebenacker},\ and\ \citenamefont
  {Svistunov}}]{Prokofev2001}%
  \BibitemOpen
  \bibfield  {author} {\bibinfo {author} {\bibfnamefont {N.}~\bibnamefont
  {Prokof'ev}}, \bibinfo {author} {\bibfnamefont {O.}~\bibnamefont
  {Ruebenacker}}, \ and\ \bibinfo {author} {\bibfnamefont {B.}~\bibnamefont
  {Svistunov}},\ }\href {\doibase 10.1103/PhysRevLett.87.270402} {\bibfield
  {journal} {\bibinfo  {journal} {Phys. Rev. Lett.}\ }\textbf {\bibinfo
  {volume} {87}},\ \bibinfo {pages} {270402} (\bibinfo {year}
  {2001})}\BibitemShut {NoStop}%
\bibitem [{\citenamefont {Prokof'ev}\ and\ \citenamefont
  {Svistunov}(2002)}]{PhysRevA.66.043608}%
  \BibitemOpen
  \bibfield  {author} {\bibinfo {author} {\bibfnamefont {N.}~\bibnamefont
  {Prokof'ev}}\ and\ \bibinfo {author} {\bibfnamefont {B.}~\bibnamefont
  {Svistunov}},\ }\href {\doibase 10.1103/PhysRevA.66.043608} {\bibfield
  {journal} {\bibinfo  {journal} {Phys. Rev. A}\ }\textbf {\bibinfo {volume}
  {66}},\ \bibinfo {pages} {043608} (\bibinfo {year} {2002})}\BibitemShut
  {NoStop}%
\bibitem [{\citenamefont {Holzmann}\ \emph {et~al.}(2007)\citenamefont
  {Holzmann}, \citenamefont {Baym}, \citenamefont {Blaizot},\ and\
  \citenamefont {Lalo\"{e}}}]{Holzmann2007}%
  \BibitemOpen
  \bibfield  {author} {\bibinfo {author} {\bibfnamefont {M.}~\bibnamefont
  {Holzmann}}, \bibinfo {author} {\bibfnamefont {G.}~\bibnamefont {Baym}},
  \bibinfo {author} {\bibfnamefont {J.-P.}\ \bibnamefont {Blaizot}}, \ and\
  \bibinfo {author} {\bibfnamefont {F.}~\bibnamefont {Lalo\"{e}}},\ }\href
  {\doibase 10.1073/pnas.0609957104} {\bibfield  {journal} {\bibinfo  {journal}
  {Proc. Natl. Acad. Sci. U.S.A.}\ }\textbf {\bibinfo {volume} {104}},\
  \bibinfo {pages} {1476} (\bibinfo {year} {2007})}\BibitemShut {NoStop}%
\bibitem [{\citenamefont {Holzmann}\ and\ \citenamefont
  {Krauth}(2008)}]{Holzmann2008}%
  \BibitemOpen
  \bibfield  {author} {\bibinfo {author} {\bibfnamefont {M.}~\bibnamefont
  {Holzmann}}\ and\ \bibinfo {author} {\bibfnamefont {W.}~\bibnamefont
  {Krauth}},\ }\href {\doibase 10.1103/PhysRevLett.100.190402} {\bibfield
  {journal} {\bibinfo  {journal} {Phys. Rev. Lett.}\ }\textbf {\bibinfo
  {volume} {100}},\ \bibinfo {pages} {190402} (\bibinfo {year}
  {2008})}\BibitemShut {NoStop}%
\bibitem [{\citenamefont {Hung}\ \emph {et~al.}(2011)\citenamefont {Hung},
  \citenamefont {Zhang}, \citenamefont {Gemelke},\ and\ \citenamefont
  {Chin}}]{Hung2011}%
  \BibitemOpen
  \bibfield  {author} {\bibinfo {author} {\bibfnamefont {C.-L.}\ \bibnamefont
  {Hung}}, \bibinfo {author} {\bibfnamefont {X.}~\bibnamefont {Zhang}},
  \bibinfo {author} {\bibfnamefont {N.}~\bibnamefont {Gemelke}}, \ and\
  \bibinfo {author} {\bibfnamefont {C.}~\bibnamefont {Chin}},\ }\href {\doibase
  10.1038/nature09722} {\bibfield  {journal} {\bibinfo  {journal} {Nature}\
  }\textbf {\bibinfo {volume} {470}},\ \bibinfo {pages} {236} (\bibinfo {year}
  {2011})}\BibitemShut {NoStop}%
\bibitem [{\citenamefont {Yefsah}\ \emph {et~al.}(2011)\citenamefont {Yefsah},
  \citenamefont {Desbuquois}, \citenamefont {Chomaz}, \citenamefont
  {G\"unter},\ and\ \citenamefont {Dalibard}}]{PhysRevLett.107.130401}%
  \BibitemOpen
  \bibfield  {author} {\bibinfo {author} {\bibfnamefont {T.}~\bibnamefont
  {Yefsah}}, \bibinfo {author} {\bibfnamefont {R.}~\bibnamefont {Desbuquois}},
  \bibinfo {author} {\bibfnamefont {L.}~\bibnamefont {Chomaz}}, \bibinfo
  {author} {\bibfnamefont {K.~J.}\ \bibnamefont {G\"unter}}, \ and\ \bibinfo
  {author} {\bibfnamefont {J.}~\bibnamefont {Dalibard}},\ }\href {\doibase
  10.1103/PhysRevLett.107.130401} {\bibfield  {journal} {\bibinfo  {journal}
  {Phys. Rev. Lett.}\ }\textbf {\bibinfo {volume} {107}},\ \bibinfo {pages}
  {130401} (\bibinfo {year} {2011})}\BibitemShut {NoStop}%
\bibitem [{\citenamefont {Zhang}\ \emph {et~al.}(2012)\citenamefont {Zhang},
  \citenamefont {Hung}, \citenamefont {Tung},\ and\ \citenamefont
  {Chin}}]{Zhang1070}%
  \BibitemOpen
  \bibfield  {author} {\bibinfo {author} {\bibfnamefont {X.}~\bibnamefont
  {Zhang}}, \bibinfo {author} {\bibfnamefont {C.-L.}\ \bibnamefont {Hung}},
  \bibinfo {author} {\bibfnamefont {S.-K.}\ \bibnamefont {Tung}}, \ and\
  \bibinfo {author} {\bibfnamefont {C.}~\bibnamefont {Chin}},\ }\href {\doibase
  10.1126/science.1217990} {\bibfield  {journal} {\bibinfo  {journal}
  {Science}\ }\textbf {\bibinfo {volume} {335}},\ \bibinfo {pages} {1070}
  (\bibinfo {year} {2012})}\BibitemShut {NoStop}%
\bibitem [{\citenamefont {Gersdorff}\ and\ \citenamefont
  {Wetterich}(2001)}]{PhysRevB.64.054513}%
  \BibitemOpen
  \bibfield  {author} {\bibinfo {author} {\bibfnamefont {G.~v.}\ \bibnamefont
  {Gersdorff}}\ and\ \bibinfo {author} {\bibfnamefont {C.}~\bibnamefont
  {Wetterich}},\ }\href {\doibase 10.1103/PhysRevB.64.054513} {\bibfield
  {journal} {\bibinfo  {journal} {Phys. Rev. B}\ }\textbf {\bibinfo {volume}
  {64}},\ \bibinfo {pages} {054513} (\bibinfo {year} {2001})}\BibitemShut
  {NoStop}%
\bibitem [{\citenamefont {Dupuis}(2009)}]{PhysRevLett.102.190401}%
  \BibitemOpen
  \bibfield  {author} {\bibinfo {author} {\bibfnamefont {N.}~\bibnamefont
  {Dupuis}},\ }\href {\doibase 10.1103/PhysRevLett.102.190401} {\bibfield
  {journal} {\bibinfo  {journal} {Phys. Rev. Lett.}\ }\textbf {\bibinfo
  {volume} {102}},\ \bibinfo {pages} {190401} (\bibinfo {year}
  {2009})}\BibitemShut {NoStop}%
\bibitem [{\citenamefont {Jakubczyk}\ \emph {et~al.}(2014)\citenamefont
  {Jakubczyk}, \citenamefont {Dupuis},\ and\ \citenamefont
  {Delamotte}}]{PhysRevE.90.062105}%
  \BibitemOpen
  \bibfield  {author} {\bibinfo {author} {\bibfnamefont {P.}~\bibnamefont
  {Jakubczyk}}, \bibinfo {author} {\bibfnamefont {N.}~\bibnamefont {Dupuis}}, \
  and\ \bibinfo {author} {\bibfnamefont {B.}~\bibnamefont {Delamotte}},\ }\href
  {\doibase 10.1103/PhysRevE.90.062105} {\bibfield  {journal} {\bibinfo
  {journal} {Phys. Rev. E}\ }\textbf {\bibinfo {volume} {90}},\ \bibinfo
  {pages} {062105} (\bibinfo {year} {2014})}\BibitemShut {NoStop}%
\bibitem [{\citenamefont {Berezinskii}(1971)}]{Berezinskii1971}%
  \BibitemOpen
  \bibfield  {author} {\bibinfo {author} {\bibfnamefont {V.~L.}\ \bibnamefont
  {Berezinskii}},\ }\href
  {http://www.jetp.ac.ru/cgi-bin/e/index/e/32/3/p493?a=list} {\bibfield
  {journal} {\bibinfo  {journal} {Sov. Phys. JETP}\ }\textbf {\bibinfo {volume}
  {32}},\ \bibinfo {pages} {493} (\bibinfo {year} {1971})}\BibitemShut
  {NoStop}%
\bibitem [{\citenamefont {Berezinskii}(1972)}]{Berezinskii1972}%
  \BibitemOpen
  \bibfield  {author} {\bibinfo {author} {\bibfnamefont {V.~L.}\ \bibnamefont
  {Berezinskii}},\ }\href
  {http://www.jetp.ac.ru/cgi-bin/index/e/34/3/p610?a=list} {\bibfield
  {journal} {\bibinfo  {journal} {Sov. Phys. JETP}\ }\textbf {\bibinfo {volume}
  {34}},\ \bibinfo {pages} {610} (\bibinfo {year} {1972})}\BibitemShut
  {NoStop}%
\bibitem [{\citenamefont {Kosterlitz}\ and\ \citenamefont
  {Thouless}(1973)}]{Kosterlitz1973}%
  \BibitemOpen
  \bibfield  {author} {\bibinfo {author} {\bibfnamefont {J.~M.}\ \bibnamefont
  {Kosterlitz}}\ and\ \bibinfo {author} {\bibfnamefont {D.~J.}\ \bibnamefont
  {Thouless}},\ }\href@noop {} {\bibfield  {journal} {\bibinfo  {journal} {J.
  Phys. C: Solid State Phys.}\ }\textbf {\bibinfo {volume} {6}},\ \bibinfo
  {pages} {1181} (\bibinfo {year} {1973})}\BibitemShut {NoStop}%
\bibitem [{\citenamefont {Kosterlitz}(1974)}]{Kosterlitz1974}%
  \BibitemOpen
  \bibfield  {author} {\bibinfo {author} {\bibfnamefont {J.~M.}\ \bibnamefont
  {Kosterlitz}},\ }\href
  {http://iopscience.iop.org/0022-3719/7/6/005$\backslash$npapers2://publication/doi/10.1088/0022-3719/7/6/005}
  {\bibfield  {journal} {\bibinfo  {journal} {J. Phys. C: Solid State Phys.}\
  }\textbf {\bibinfo {volume} {7}},\ \bibinfo {pages} {1046} (\bibinfo {year}
  {1974})}\BibitemShut {NoStop}%
\bibitem [{\citenamefont {Al~Khawaja}\ \emph {et~al.}(2003)\citenamefont
  {Al~Khawaja}, \citenamefont {Proukakis}, \citenamefont {Andersen},
  \citenamefont {Romans},\ and\ \citenamefont {Stoof}}]{PhysRevA.68.043603}%
  \BibitemOpen
  \bibfield  {author} {\bibinfo {author} {\bibfnamefont {U.}~\bibnamefont
  {Al~Khawaja}}, \bibinfo {author} {\bibfnamefont {N.~P.}\ \bibnamefont
  {Proukakis}}, \bibinfo {author} {\bibfnamefont {J.~O.}\ \bibnamefont
  {Andersen}}, \bibinfo {author} {\bibfnamefont {M.~W.~J.}\ \bibnamefont
  {Romans}}, \ and\ \bibinfo {author} {\bibfnamefont {H.~T.~C.}\ \bibnamefont
  {Stoof}},\ }\href {\doibase 10.1103/PhysRevA.68.043603} {\bibfield  {journal}
  {\bibinfo  {journal} {Phys. Rev. A}\ }\textbf {\bibinfo {volume} {68}},\
  \bibinfo {pages} {043603} (\bibinfo {year} {2003})}\BibitemShut {NoStop}%
\bibitem [{\citenamefont {{Braun}}\ and\ \citenamefont
  {{Klein}}(2009)}]{2009EPJC...63..443B}%
  \BibitemOpen
  \bibfield  {author} {\bibinfo {author} {\bibfnamefont {J.}~\bibnamefont
  {{Braun}}}\ and\ \bibinfo {author} {\bibfnamefont {B.}~\bibnamefont
  {{Klein}}},\ }\href {\doibase 10.1140/epjc/s10052-009-1098-8} {\bibfield
  {journal} {\bibinfo  {journal} {European Physical Journal C}\ }\textbf
  {\bibinfo {volume} {63}},\ \bibinfo {pages} {443} (\bibinfo {year}
  {2009})}\BibitemShut {NoStop}%
\bibitem [{\citenamefont {{Braun}}\ \emph
  {et~al.}(2011{\natexlab{a}})\citenamefont {{Braun}}, \citenamefont
  {{Klein}},\ and\ \citenamefont {{Piasecki}}}]{2011EPJC...71.1576B}%
  \BibitemOpen
  \bibfield  {author} {\bibinfo {author} {\bibfnamefont {J.}~\bibnamefont
  {{Braun}}}, \bibinfo {author} {\bibfnamefont {B.}~\bibnamefont {{Klein}}}, \
  and\ \bibinfo {author} {\bibfnamefont {P.}~\bibnamefont {{Piasecki}}},\
  }\href {\doibase 10.1140/epjc/s10052-011-1576-7} {\bibfield  {journal}
  {\bibinfo  {journal} {European Physical Journal C}\ }\textbf {\bibinfo
  {volume} {71}},\ \bibinfo {eid} {1576} (\bibinfo {year}
  {2011}{\natexlab{a}})}\BibitemShut {NoStop}%
\bibitem [{\citenamefont {{Braun}}\ \emph {et~al.}(2012)\citenamefont
  {{Braun}}, \citenamefont {{Klein}},\ and\ \citenamefont
  {{Schaefer}}}]{2012PhLB..713..216B}%
  \BibitemOpen
  \bibfield  {author} {\bibinfo {author} {\bibfnamefont {J.}~\bibnamefont
  {{Braun}}}, \bibinfo {author} {\bibfnamefont {B.}~\bibnamefont {{Klein}}}, \
  and\ \bibinfo {author} {\bibfnamefont {B.-J.}\ \bibnamefont {{Schaefer}}},\
  }\href {\doibase 10.1016/j.physletb.2012.05.053} {\bibfield  {journal}
  {\bibinfo  {journal} {Physics Letters B}\ }\textbf {\bibinfo {volume}
  {713}},\ \bibinfo {pages} {216} (\bibinfo {year} {2012})}\BibitemShut
  {NoStop}%
\bibitem [{\citenamefont {{Tripolt}}\ \emph {et~al.}(2014)\citenamefont
  {{Tripolt}}, \citenamefont {{Braun}}, \citenamefont {{Klein}},\ and\
  \citenamefont {{Schaefer}}}]{2014PhRvD..90e4012T}%
  \BibitemOpen
  \bibfield  {author} {\bibinfo {author} {\bibfnamefont {R.-A.}\ \bibnamefont
  {{Tripolt}}}, \bibinfo {author} {\bibfnamefont {J.}~\bibnamefont {{Braun}}},
  \bibinfo {author} {\bibfnamefont {B.}~\bibnamefont {{Klein}}}, \ and\
  \bibinfo {author} {\bibfnamefont {B.-J.}\ \bibnamefont {{Schaefer}}},\ }\href
  {\doibase 10.1103/PhysRevD.90.054012} {\bibfield  {journal} {\bibinfo
  {journal} {\prd}\ }\textbf {\bibinfo {volume} {90}},\ \bibinfo {eid} {054012}
  (\bibinfo {year} {2014})}\BibitemShut {NoStop}%
\bibitem [{\citenamefont {{Braun}}\ \emph
  {et~al.}(2011{\natexlab{b}})\citenamefont {{Braun}}, \citenamefont
  {{Diehl}},\ and\ \citenamefont {{Scherer}}}]{2011PhRvA..84f3616B}%
  \BibitemOpen
  \bibfield  {author} {\bibinfo {author} {\bibfnamefont {J.}~\bibnamefont
  {{Braun}}}, \bibinfo {author} {\bibfnamefont {S.}~\bibnamefont {{Diehl}}}, \
  and\ \bibinfo {author} {\bibfnamefont {M.~M.}\ \bibnamefont {{Scherer}}},\
  }\href {\doibase 10.1103/PhysRevA.84.063616} {\bibfield  {journal} {\bibinfo
  {journal} {\pra}\ }\textbf {\bibinfo {volume} {84}},\ \bibinfo {eid} {063616}
  (\bibinfo {year} {2011}{\natexlab{b}})}\BibitemShut {NoStop}%
\bibitem [{\citenamefont {Tetradis}\ and\ \citenamefont
  {Wetterich}(1993)}]{Tetradis:1992xd}%
  \BibitemOpen
  \bibfield  {author} {\bibinfo {author} {\bibfnamefont {N.}~\bibnamefont
  {Tetradis}}\ and\ \bibinfo {author} {\bibfnamefont {C.}~\bibnamefont
  {Wetterich}},\ }\href {\doibase 10.1016/0550-3213(93)90608-R} {\bibfield
  {journal} {\bibinfo  {journal} {Nucl. Phys.}\ }\textbf {\bibinfo {volume}
  {B398}},\ \bibinfo {pages} {659} (\bibinfo {year} {1993})}\BibitemShut
  {NoStop}%
\bibitem [{\citenamefont {Tetradis}\ and\ \citenamefont
  {Wetterich}(1994{\natexlab{b}})}]{Tetradis:1993bx}%
  \BibitemOpen
  \bibfield  {author} {\bibinfo {author} {\bibfnamefont {N.}~\bibnamefont
  {Tetradis}}\ and\ \bibinfo {author} {\bibfnamefont {C.}~\bibnamefont
  {Wetterich}},\ }\href {\doibase 10.1142/S0217751X94001631} {\bibfield
  {journal} {\bibinfo  {journal} {Int. J. Mod. Phys.}\ }\textbf {\bibinfo
  {volume} {A9}},\ \bibinfo {pages} {4029} (\bibinfo {year}
  {1994}{\natexlab{b}})}\BibitemShut {NoStop}%
\bibitem [{\citenamefont {Gr\"uter}\ \emph {et~al.}(1997)\citenamefont
  {Gr\"uter}, \citenamefont {Ceperley},\ and\ \citenamefont
  {Lalo\"e}}]{PhysRevLett.79.3549}%
  \BibitemOpen
  \bibfield  {author} {\bibinfo {author} {\bibfnamefont {P.}~\bibnamefont
  {Gr\"uter}}, \bibinfo {author} {\bibfnamefont {D.}~\bibnamefont {Ceperley}},
  \ and\ \bibinfo {author} {\bibfnamefont {F.}~\bibnamefont {Lalo\"e}},\ }\href
  {\doibase 10.1103/PhysRevLett.79.3549} {\bibfield  {journal} {\bibinfo
  {journal} {Phys. Rev. Lett.}\ }\textbf {\bibinfo {volume} {79}},\ \bibinfo
  {pages} {3549} (\bibinfo {year} {1997})}\BibitemShut {NoStop}%
\bibitem [{\citenamefont {Baym}\ \emph {et~al.}(1999)\citenamefont {Baym},
  \citenamefont {Blaizot}, \citenamefont {Holzmann}, \citenamefont {Lalo\"e},\
  and\ \citenamefont {Vautherin}}]{PhysRevLett.83.1703}%
  \BibitemOpen
  \bibfield  {author} {\bibinfo {author} {\bibfnamefont {G.}~\bibnamefont
  {Baym}}, \bibinfo {author} {\bibfnamefont {J.-P.}\ \bibnamefont {Blaizot}},
  \bibinfo {author} {\bibfnamefont {M.}~\bibnamefont {Holzmann}}, \bibinfo
  {author} {\bibfnamefont {F.}~\bibnamefont {Lalo\"e}}, \ and\ \bibinfo
  {author} {\bibfnamefont {D.}~\bibnamefont {Vautherin}},\ }\href {\doibase
  10.1103/PhysRevLett.83.1703} {\bibfield  {journal} {\bibinfo  {journal}
  {Phys. Rev. Lett.}\ }\textbf {\bibinfo {volume} {83}},\ \bibinfo {pages}
  {1703} (\bibinfo {year} {1999})}\BibitemShut {NoStop}%
\bibitem [{\citenamefont {Holzmann}\ and\ \citenamefont
  {Krauth}(1999)}]{PhysRevLett.83.2687}%
  \BibitemOpen
  \bibfield  {author} {\bibinfo {author} {\bibfnamefont {M.}~\bibnamefont
  {Holzmann}}\ and\ \bibinfo {author} {\bibfnamefont {W.}~\bibnamefont
  {Krauth}},\ }\href {\doibase 10.1103/PhysRevLett.83.2687} {\bibfield
  {journal} {\bibinfo  {journal} {Phys. Rev. Lett.}\ }\textbf {\bibinfo
  {volume} {83}},\ \bibinfo {pages} {2687} (\bibinfo {year}
  {1999})}\BibitemShut {NoStop}%
\bibitem [{\citenamefont {Arnold}\ and\ \citenamefont
  {Moore}(2001)}]{PhysRevLett.87.120401}%
  \BibitemOpen
  \bibfield  {author} {\bibinfo {author} {\bibfnamefont {P.}~\bibnamefont
  {Arnold}}\ and\ \bibinfo {author} {\bibfnamefont {G.}~\bibnamefont {Moore}},\
  }\href {\doibase 10.1103/PhysRevLett.87.120401} {\bibfield  {journal}
  {\bibinfo  {journal} {Phys. Rev. Lett.}\ }\textbf {\bibinfo {volume} {87}},\
  \bibinfo {pages} {120401} (\bibinfo {year} {2001})}\BibitemShut {NoStop}%
\bibitem [{\citenamefont {Kashurnikov}\ \emph {et~al.}(2001)\citenamefont
  {Kashurnikov}, \citenamefont {Prokof'ev},\ and\ \citenamefont
  {Svistunov}}]{PhysRevLett.87.120402}%
  \BibitemOpen
  \bibfield  {author} {\bibinfo {author} {\bibfnamefont {V.~A.}\ \bibnamefont
  {Kashurnikov}}, \bibinfo {author} {\bibfnamefont {N.~V.}\ \bibnamefont
  {Prokof'ev}}, \ and\ \bibinfo {author} {\bibfnamefont {B.~V.}\ \bibnamefont
  {Svistunov}},\ }\href {\doibase 10.1103/PhysRevLett.87.120402} {\bibfield
  {journal} {\bibinfo  {journal} {Phys. Rev. Lett.}\ }\textbf {\bibinfo
  {volume} {87}},\ \bibinfo {pages} {120402} (\bibinfo {year}
  {2001})}\BibitemShut {NoStop}%
\bibitem [{\citenamefont {Fischer}\ and\ \citenamefont
  {Parish}(2014)}]{PhysRevB.90.214503}%
  \BibitemOpen
  \bibfield  {author} {\bibinfo {author} {\bibfnamefont {A.~M.}\ \bibnamefont
  {Fischer}}\ and\ \bibinfo {author} {\bibfnamefont {M.~M.}\ \bibnamefont
  {Parish}},\ }\href {\doibase 10.1103/PhysRevB.90.214503} {\bibfield
  {journal} {\bibinfo  {journal} {Phys. Rev. B}\ }\textbf {\bibinfo {volume}
  {90}},\ \bibinfo {pages} {214503} (\bibinfo {year} {2014})}\BibitemShut
  {NoStop}%
\bibitem [{\citenamefont {Petrov}\ and\ \citenamefont
  {Shlyapnikov}(2001)}]{PhysRevA.64.012706}%
  \BibitemOpen
  \bibfield  {author} {\bibinfo {author} {\bibfnamefont {D.~S.}\ \bibnamefont
  {Petrov}}\ and\ \bibinfo {author} {\bibfnamefont {G.~V.}\ \bibnamefont
  {Shlyapnikov}},\ }\href {\doibase 10.1103/PhysRevA.64.012706} {\bibfield
  {journal} {\bibinfo  {journal} {Phys. Rev. A}\ }\textbf {\bibinfo {volume}
  {64}},\ \bibinfo {pages} {012706} (\bibinfo {year} {2001})}\BibitemShut
  {NoStop}%
\bibitem [{\citenamefont {Levinsen}\ and\ \citenamefont
  {Parish}(2015)}]{Levinsen2015}%
  \BibitemOpen
  \bibfield  {author} {\bibinfo {author} {\bibfnamefont {J.}~\bibnamefont
  {Levinsen}}\ and\ \bibinfo {author} {\bibfnamefont {M.~M.}\ \bibnamefont
  {Parish}},\ }\href
  {http://www.worldscientific.com/doi/abs/10.1142/9789814667746_0001}
  {\bibfield  {journal} {\bibinfo  {journal} {Annual Review of Cold Atoms and
  Molecules}\ }\textbf {\bibinfo {volume} {3}},\ \bibinfo {pages} {1} (\bibinfo
  {year} {2015})}\BibitemShut {NoStop}%
\bibitem [{\citenamefont {{Floerchinger}}(2014)}]{Floerchinger:2013tp}%
  \BibitemOpen
  \bibfield  {author} {\bibinfo {author} {\bibfnamefont {S.}~\bibnamefont
  {{Floerchinger}}},\ }\href {\doibase 10.1016/j.nuclphysa.2014.04.013}
  {\bibfield  {journal} {\bibinfo  {journal} {Nuclear Physics A}\ }\textbf
  {\bibinfo {volume} {927}},\ \bibinfo {pages} {119} (\bibinfo {year}
  {2014})}\BibitemShut {NoStop}%
\bibitem [{\citenamefont {{Yamashita}}\ \emph {et~al.}(2015)\citenamefont
  {{Yamashita}}, \citenamefont {{Bellotti}}, \citenamefont {{Frederico}},
  \citenamefont {{Fedorov}}, \citenamefont {{Jensen}},\ and\ \citenamefont
  {{Zinner}}}]{2015JPhB...48b5302Y}%
  \BibitemOpen
  \bibfield  {author} {\bibinfo {author} {\bibfnamefont {M.~T.}\ \bibnamefont
  {{Yamashita}}}, \bibinfo {author} {\bibfnamefont {F.~F.}\ \bibnamefont
  {{Bellotti}}}, \bibinfo {author} {\bibfnamefont {T.}~\bibnamefont
  {{Frederico}}}, \bibinfo {author} {\bibfnamefont {D.~V.}\ \bibnamefont
  {{Fedorov}}}, \bibinfo {author} {\bibfnamefont {A.~S.}\ \bibnamefont
  {{Jensen}}}, \ and\ \bibinfo {author} {\bibfnamefont {N.~T.}\ \bibnamefont
  {{Zinner}}},\ }\href {\doibase 10.1088/0953-4075/48/2/025302} {\bibfield
  {journal} {\bibinfo  {journal} {Journal of Physics B Atomic Molecular
  Physics}\ }\textbf {\bibinfo {volume} {48}},\ \bibinfo {eid} {025302}
  (\bibinfo {year} {2015})},\ \Eprint {http://arxiv.org/abs/1404.7002}
  {arXiv:1404.7002 [cond-mat.quant-gas]} \BibitemShut {NoStop}%
\bibitem [{\citenamefont {Morgan}\ \emph {et~al.}(2002)\citenamefont {Morgan},
  \citenamefont {Lee},\ and\ \citenamefont {Burnett}}]{PhysRevA.65.022706}%
  \BibitemOpen
  \bibfield  {author} {\bibinfo {author} {\bibfnamefont {S.~A.}\ \bibnamefont
  {Morgan}}, \bibinfo {author} {\bibfnamefont {M.~D.}\ \bibnamefont {Lee}}, \
  and\ \bibinfo {author} {\bibfnamefont {K.}~\bibnamefont {Burnett}},\ }\href
  {\doibase 10.1103/PhysRevA.65.022706} {\bibfield  {journal} {\bibinfo
  {journal} {Phys. Rev. A}\ }\textbf {\bibinfo {volume} {65}},\ \bibinfo
  {pages} {022706} (\bibinfo {year} {2002})}\BibitemShut {NoStop}%
\bibitem [{\citenamefont {Sun}\ and\ \citenamefont
  {Bolech}(2012)}]{PhysRevA.85.051607}%
  \BibitemOpen
  \bibfield  {author} {\bibinfo {author} {\bibfnamefont {K.}~\bibnamefont
  {Sun}}\ and\ \bibinfo {author} {\bibfnamefont {C.~J.}\ \bibnamefont
  {Bolech}},\ }\href {\doibase 10.1103/PhysRevA.85.051607} {\bibfield
  {journal} {\bibinfo  {journal} {Phys. Rev. A}\ }\textbf {\bibinfo {volume}
  {85}},\ \bibinfo {pages} {051607} (\bibinfo {year} {2012})}\BibitemShut
  {NoStop}%
\bibitem [{\citenamefont {{Dutta}}\ and\ \citenamefont
  {{Mueller}}(2015)}]{2015arXiv150803352D}%
  \BibitemOpen
  \bibfield  {author} {\bibinfo {author} {\bibfnamefont {S.}~\bibnamefont
  {{Dutta}}}\ and\ \bibinfo {author} {\bibfnamefont {E.~J.}\ \bibnamefont
  {{Mueller}}},\ }\href@noop {} {\  (\bibinfo {year} {2015})},\ \Eprint
  {http://arxiv.org/abs/1508.03352} {arXiv:1508.03352} \BibitemShut {NoStop}%
\bibitem [{\citenamefont {Lee}\ \emph {et~al.}(2006)\citenamefont {Lee},
  \citenamefont {Nagaosa},\ and\ \citenamefont {Wen}}]{RevModPhys.78.17}%
  \BibitemOpen
  \bibfield  {author} {\bibinfo {author} {\bibfnamefont {P.~A.}\ \bibnamefont
  {Lee}}, \bibinfo {author} {\bibfnamefont {N.}~\bibnamefont {Nagaosa}}, \ and\
  \bibinfo {author} {\bibfnamefont {X.-G.}\ \bibnamefont {Wen}},\ }\href
  {\doibase 10.1103/RevModPhys.78.17} {\bibfield  {journal} {\bibinfo
  {journal} {Rev. Mod. Phys.}\ }\textbf {\bibinfo {volume} {78}},\ \bibinfo
  {pages} {17} (\bibinfo {year} {2006})}\BibitemShut {NoStop}%
\bibitem [{\citenamefont {Carusotto}\ and\ \citenamefont
  {Ciuti}(2013)}]{RevModPhys.85.299}%
  \BibitemOpen
  \bibfield  {author} {\bibinfo {author} {\bibfnamefont {I.}~\bibnamefont
  {Carusotto}}\ and\ \bibinfo {author} {\bibfnamefont {C.}~\bibnamefont
  {Ciuti}},\ }\href {\doibase 10.1103/RevModPhys.85.299} {\bibfield  {journal}
  {\bibinfo  {journal} {Rev. Mod. Phys.}\ }\textbf {\bibinfo {volume} {85}},\
  \bibinfo {pages} {299} (\bibinfo {year} {2013})}\BibitemShut {NoStop}%
\bibitem [{\citenamefont {{Sieberer}}\ \emph {et~al.}(2015)\citenamefont
  {{Sieberer}}, \citenamefont {{Buchhold}},\ and\ \citenamefont
  {{Diehl}}}]{2015arXiv151200637S}%
  \BibitemOpen
  \bibfield  {author} {\bibinfo {author} {\bibfnamefont {L.~M.}\ \bibnamefont
  {{Sieberer}}}, \bibinfo {author} {\bibfnamefont {M.}~\bibnamefont
  {{Buchhold}}}, \ and\ \bibinfo {author} {\bibfnamefont {S.}~\bibnamefont
  {{Diehl}}},\ }\href@noop {} {\bibfield  {journal} {\bibinfo  {journal} {ArXiv
  e-prints}\ } (\bibinfo {year} {2015})},\ \Eprint
  {http://arxiv.org/abs/1512.00637} {arXiv:1512.00637 [cond-mat.quant-gas]}
  \BibitemShut {NoStop}%
\bibitem [{\citenamefont {Berges}\ and\ \citenamefont
  {Mesterhazy}(2012)}]{Berges:2012ty}%
  \BibitemOpen
  \bibfield  {author} {\bibinfo {author} {\bibfnamefont {J.}~\bibnamefont
  {Berges}}\ and\ \bibinfo {author} {\bibfnamefont {D.}~\bibnamefont
  {Mesterhazy}},\ }\bibfield  {booktitle} {\emph {\bibinfo {booktitle}
  {{Physics at all scales: The Renormalization Group. Proceedings, 49.
  Internationale Universitätswochen für Theoretische Physik, Winter
  School}}},\ }\href {\doibase 10.1016/j.nuclphysbps.2012.06.003} {\bibfield
  {journal} {\bibinfo  {journal} {Nucl. Phys. Proc. Suppl.}\ }\textbf {\bibinfo
  {volume} {228}},\ \bibinfo {pages} {37} (\bibinfo {year} {2012})}\BibitemShut
  {NoStop}%
\bibitem [{\citenamefont {Makhalov}\ \emph {et~al.}(2014)\citenamefont
  {Makhalov}, \citenamefont {Martiyanov},\ and\ \citenamefont
  {Turlapov}}]{PhysRevLett.112.045301}%
  \BibitemOpen
  \bibfield  {author} {\bibinfo {author} {\bibfnamefont {V.}~\bibnamefont
  {Makhalov}}, \bibinfo {author} {\bibfnamefont {K.}~\bibnamefont
  {Martiyanov}}, \ and\ \bibinfo {author} {\bibfnamefont {A.}~\bibnamefont
  {Turlapov}},\ }\href {\doibase 10.1103/PhysRevLett.112.045301} {\bibfield
  {journal} {\bibinfo  {journal} {Phys. Rev. Lett.}\ }\textbf {\bibinfo
  {volume} {112}},\ \bibinfo {pages} {045301} (\bibinfo {year}
  {2014})}\BibitemShut {NoStop}%
\bibitem [{\citenamefont {Dyke}\ \emph {et~al.}(2016)\citenamefont {Dyke},
  \citenamefont {Fenech}, \citenamefont {Peppler}, \citenamefont {Lingham},
  \citenamefont {Hoinka}, \citenamefont {Zhang}, \citenamefont {Peng},
  \citenamefont {Mulkerin}, \citenamefont {Hu}, \citenamefont {Liu},\ and\
  \citenamefont {Vale}}]{Dyke2014}%
  \BibitemOpen
  \bibfield  {author} {\bibinfo {author} {\bibfnamefont {P.}~\bibnamefont
  {Dyke}}, \bibinfo {author} {\bibfnamefont {K.}~\bibnamefont {Fenech}},
  \bibinfo {author} {\bibfnamefont {T.}~\bibnamefont {Peppler}}, \bibinfo
  {author} {\bibfnamefont {M.~G.}\ \bibnamefont {Lingham}}, \bibinfo {author}
  {\bibfnamefont {S.}~\bibnamefont {Hoinka}}, \bibinfo {author} {\bibfnamefont
  {W.}~\bibnamefont {Zhang}}, \bibinfo {author} {\bibfnamefont {S.-G.}\
  \bibnamefont {Peng}}, \bibinfo {author} {\bibfnamefont {B.}~\bibnamefont
  {Mulkerin}}, \bibinfo {author} {\bibfnamefont {H.}~\bibnamefont {Hu}},
  \bibinfo {author} {\bibfnamefont {X.-J.}\ \bibnamefont {Liu}}, \ and\
  \bibinfo {author} {\bibfnamefont {C.~J.}\ \bibnamefont {Vale}},\ }\href
  {\doibase 10.1103/PhysRevA.93.011603} {\bibfield  {journal} {\bibinfo
  {journal} {Phys. Rev. A}\ }\textbf {\bibinfo {volume} {93}},\ \bibinfo
  {pages} {011603} (\bibinfo {year} {2016})}\BibitemShut {NoStop}%
\bibitem [{\citenamefont {Fenech}\ \emph {et~al.}(2016)\citenamefont {Fenech},
  \citenamefont {Dyke}, \citenamefont {Peppler}, \citenamefont {Lingham},
  \citenamefont {Hoinka}, \citenamefont {Hu},\ and\ \citenamefont
  {Vale}}]{Fenech2015}%
  \BibitemOpen
  \bibfield  {author} {\bibinfo {author} {\bibfnamefont {K.}~\bibnamefont
  {Fenech}}, \bibinfo {author} {\bibfnamefont {P.}~\bibnamefont {Dyke}},
  \bibinfo {author} {\bibfnamefont {T.}~\bibnamefont {Peppler}}, \bibinfo
  {author} {\bibfnamefont {M.~G.}\ \bibnamefont {Lingham}}, \bibinfo {author}
  {\bibfnamefont {S.}~\bibnamefont {Hoinka}}, \bibinfo {author} {\bibfnamefont
  {H.}~\bibnamefont {Hu}}, \ and\ \bibinfo {author} {\bibfnamefont {C.~J.}\
  \bibnamefont {Vale}},\ }\href {\doibase 10.1103/PhysRevLett.116.045302}
  {\bibfield  {journal} {\bibinfo  {journal} {Phys. Rev. Lett.}\ }\textbf
  {\bibinfo {volume} {116}},\ \bibinfo {pages} {045302} (\bibinfo {year}
  {2016})}\BibitemShut {NoStop}%
\bibitem [{\citenamefont {Boettcher}\ \emph {et~al.}(2016)\citenamefont
  {Boettcher}, \citenamefont {Bayha}, \citenamefont {Kedar}, \citenamefont
  {Murthy}, \citenamefont {Neidig}, \citenamefont {Ries}, \citenamefont {Wenz},
  \citenamefont {Z\"urn}, \citenamefont {Jochim},\ and\ \citenamefont
  {Enss}}]{Boettcher2015}%
  \BibitemOpen
  \bibfield  {author} {\bibinfo {author} {\bibfnamefont {I.}~\bibnamefont
  {Boettcher}}, \bibinfo {author} {\bibfnamefont {L.}~\bibnamefont {Bayha}},
  \bibinfo {author} {\bibfnamefont {D.}~\bibnamefont {Kedar}}, \bibinfo
  {author} {\bibfnamefont {P.~A.}\ \bibnamefont {Murthy}}, \bibinfo {author}
  {\bibfnamefont {M.}~\bibnamefont {Neidig}}, \bibinfo {author} {\bibfnamefont
  {M.~G.}\ \bibnamefont {Ries}}, \bibinfo {author} {\bibfnamefont {A.~N.}\
  \bibnamefont {Wenz}}, \bibinfo {author} {\bibfnamefont {G.}~\bibnamefont
  {Z\"urn}}, \bibinfo {author} {\bibfnamefont {S.}~\bibnamefont {Jochim}}, \
  and\ \bibinfo {author} {\bibfnamefont {T.}~\bibnamefont {Enss}},\ }\href
  {\doibase 10.1103/PhysRevLett.116.045303} {\bibfield  {journal} {\bibinfo
  {journal} {Phys. Rev. Lett.}\ }\textbf {\bibinfo {volume} {116}},\ \bibinfo
  {pages} {045303} (\bibinfo {year} {2016})}\BibitemShut {NoStop}%
\bibitem [{\citenamefont {Fuchs}\ \emph {et~al.}(2003)\citenamefont {Fuchs},
  \citenamefont {Leyronas},\ and\ \citenamefont
  {Combescot}}]{PhysRevA.68.043610}%
  \BibitemOpen
  \bibfield  {author} {\bibinfo {author} {\bibfnamefont {J.~N.}\ \bibnamefont
  {Fuchs}}, \bibinfo {author} {\bibfnamefont {X.}~\bibnamefont {Leyronas}}, \
  and\ \bibinfo {author} {\bibfnamefont {R.}~\bibnamefont {Combescot}},\ }\href
  {\doibase 10.1103/PhysRevA.68.043610} {\bibfield  {journal} {\bibinfo
  {journal} {Phys. Rev. A}\ }\textbf {\bibinfo {volume} {68}},\ \bibinfo
  {pages} {043610} (\bibinfo {year} {2003})}\BibitemShut {NoStop}%
\bibitem [{\citenamefont {Heiselberg}(2004)}]{PhysRevLett.93.040402}%
  \BibitemOpen
  \bibfield  {author} {\bibinfo {author} {\bibfnamefont {H.}~\bibnamefont
  {Heiselberg}},\ }\href {\doibase 10.1103/PhysRevLett.93.040402} {\bibfield
  {journal} {\bibinfo  {journal} {Phys. Rev. Lett.}\ }\textbf {\bibinfo
  {volume} {93}},\ \bibinfo {pages} {040402} (\bibinfo {year}
  {2004})}\BibitemShut {NoStop}%
\bibitem [{\citenamefont {Boettcher}\ \emph {et~al.}(2011)\citenamefont
  {Boettcher}, \citenamefont {Floerchinger},\ and\ \citenamefont
  {Wetterich}}]{Boettcher:2011iq}%
  \BibitemOpen
  \bibfield  {author} {\bibinfo {author} {\bibfnamefont {I.}~\bibnamefont
  {Boettcher}}, \bibinfo {author} {\bibfnamefont {S.}~\bibnamefont
  {Floerchinger}}, \ and\ \bibinfo {author} {\bibfnamefont {C.}~\bibnamefont
  {Wetterich}},\ }\href {\doibase 10.1088/0953-4075/44/23/235301} {\bibfield
  {journal} {\bibinfo  {journal} {J. Phys.}\ }\textbf {\bibinfo {volume}
  {B44}},\ \bibinfo {pages} {235301} (\bibinfo {year} {2011})}\BibitemShut
  {NoStop}%
\bibitem [{\citenamefont {Schmidt}\ \emph {et~al.}(2012)\citenamefont
  {Schmidt}, \citenamefont {Enss}, \citenamefont {Pietil\"a},\ and\
  \citenamefont {Demler}}]{PhysRevA.85.021602}%
  \BibitemOpen
  \bibfield  {author} {\bibinfo {author} {\bibfnamefont {R.}~\bibnamefont
  {Schmidt}}, \bibinfo {author} {\bibfnamefont {T.}~\bibnamefont {Enss}},
  \bibinfo {author} {\bibfnamefont {V.}~\bibnamefont {Pietil\"a}}, \ and\
  \bibinfo {author} {\bibfnamefont {E.}~\bibnamefont {Demler}},\ }\href
  {\doibase 10.1103/PhysRevA.85.021602} {\bibfield  {journal} {\bibinfo
  {journal} {Phys. Rev. A}\ }\textbf {\bibinfo {volume} {85}},\ \bibinfo
  {pages} {021602} (\bibinfo {year} {2012})}\BibitemShut {NoStop}%
\bibitem [{\citenamefont {Boettcher}\ \emph {et~al.}(2014)\citenamefont
  {Boettcher}, \citenamefont {Pawlowski},\ and\ \citenamefont
  {Wetterich}}]{PhysRevA.89.053630}%
  \BibitemOpen
  \bibfield  {author} {\bibinfo {author} {\bibfnamefont {I.}~\bibnamefont
  {Boettcher}}, \bibinfo {author} {\bibfnamefont {J.~M.}\ \bibnamefont
  {Pawlowski}}, \ and\ \bibinfo {author} {\bibfnamefont {C.}~\bibnamefont
  {Wetterich}},\ }\href {\doibase 10.1103/PhysRevA.89.053630} {\bibfield
  {journal} {\bibinfo  {journal} {Phys. Rev. A}\ }\textbf {\bibinfo {volume}
  {89}},\ \bibinfo {pages} {053630} (\bibinfo {year} {2014})}\BibitemShut
  {NoStop}%
\bibitem [{\citenamefont {Litim}(2001)}]{Litim:2001fd}%
  \BibitemOpen
  \bibfield  {author} {\bibinfo {author} {\bibfnamefont {D.~F.}\ \bibnamefont
  {Litim}},\ }\href {\doibase 10.1142/S0217751X01004748} {\bibfield  {journal}
  {\bibinfo  {journal} {Int.J.Mod.Phys.}\ }\textbf {\bibinfo {volume} {A16}},\
  \bibinfo {pages} {2081} (\bibinfo {year} {2001})}\BibitemShut {NoStop}%
\bibitem [{\citenamefont {Litim}(2000)}]{Litim:2000ci}%
  \BibitemOpen
  \bibfield  {author} {\bibinfo {author} {\bibfnamefont {D.~F.}\ \bibnamefont
  {Litim}},\ }\href {\doibase 10.1016/S0370-2693(00)00748-6} {\bibfield
  {journal} {\bibinfo  {journal} {Phys.Lett.}\ }\textbf {\bibinfo {volume}
  {B486}},\ \bibinfo {pages} {92} (\bibinfo {year} {2000})}\BibitemShut
  {NoStop}%
\end{thebibliography}%

\end{document}